\documentclass[a4paper,preprint,superscriptaddress,preprintnumbers,nofootinbib]{revtex4}

\usepackage{mathrsfs}
\usepackage{amsmath,bm,amssymb}
\usepackage{graphicx}

\newcommand{\Slash}[1]{{\ooalign{\hfil/\hfil\crcr$#1$}}}

\long\def\CFF#1#2{
\hbox to \hsize{\hfill\includegraphics[width=#1mm,clip]{#2.pdf}\hfill}
}
\long\def\FF#1#2{
\hbox to #1mm{\hfill\includegraphics[width=#1mm,clip]{#2.pdf}\hfill}
}
\def\cal#1{\mathcal{#1}}

\begin{document}

\renewcommand{\thefootnote}{\fnsymbol{footnote}}
\footnote[0]{To be published in {\em The Science Reports of Kanazawa University} {\bf 61} (2017), 1.}
\renewcommand{\thefootnote}{\arabic{footnote}}

\title{Singularity Free Direct Calculation of \\Spontaneous Mass Generation}

\author{Ken-Ichi \surname{Aoki}}
\email{aoki@hep.s.kanazawa-u.ac.jp}
\affiliation{Institute for Theoretical Physics, Kanazawa University, Kanazawa 920-1192, Japan}

\author{Tamao \surname{Kobayashi }}
\email{kobayasi@yonago-k.ac.jp}
\affiliation{Yonago College of Technology, Yonago 683-8502, Japan}

\author{Shin-Ichiro \surname{Kumamoto}}
\email{kumamoto@rieb.kobe-u.ac.jp}
\affiliation{Research Institute for Economics and Business Administration, Kobe University, Kobe 657-8501, Japan}

\author{Shinnosuke \surname{Onai}}
\email{onai@hep.s.kanazawa-u.ac.jp}
\affiliation{Institute for Theoretical Physics, Kanazawa University, Kanazawa 920-1192, Japan}

\author{Daisuke \surname{Sato}\footnote{until September 2014}}
\email{satodai@hep.s.kanazawa-u.ac.jp}
\affiliation{Institute for Theoretical Physics, Kanazawa University, Kanazawa 920-1192, Japan}

\preprint{KANAZAWA-18-08}

\begin{abstract}
We propose a new iterative method to directly calculate
the spontaneous mass generation. It is regarded as a new regularization
method resembling the finite volume calculation 
which assures non-negative fluctuation property at every stage.
We work with the Nambu--Jona-Lasinio model 
and the strong coupling gauge theory where the dynamical
chiral symmetry breaking occurs. 
We are able to conclude the physical mass definitely without
encountering any singularity nor 
recourse to any additional consideration like the free energy comparison.
However in special case of the 1st order phase transition, we find that
the iterative method has a chance to go wrong.
\end{abstract}

\maketitle

\section{Introduction}

Owing to its outstanding feature that it is the central issue of the elementary
particle physics, dynamical chiral symmetry breaking phenomena
have been widely studied using various methods. The
standard method to discuss the spontaneous mass generation is to formulate a
coupled system of self-consistent equations and find its non-trivial solution. 
However, those equations are no more than the {\sl necessary} condition and it is
needed to examine solutions to select correct one by using another means, e.g., by
referring to the free energy of each solution. Even if it is done, it is 
still not completely clear whether the minimal free energy solution ensures the
physically correct answer.

The method of setting up self-consistent equations for some infinite summation
is based on the observation
that the whole (T) resides in the whole as a part P. There must be a function $f$ that
the whole is calculated by using the part, 
\begin{equation}
T=f(P).
\end{equation}
Then we have a self-consistent equation,
\begin{equation}
T=f(T),
\end{equation}
and try to find solutions of this equation.
Usually there are many solutions and we must proceed to pick up
the physically correct one, supposing it is there anyway.

In contrast to the self-consistent equation method, we propose another
method of iteration. We define a series of parts $P^{(n)}$, numbered by $n$,
 so that it has a feature:
\begin{equation}
\lim_{n\rightarrow\infty}P^{(n)} = T.
\end{equation}
If we find an iterative relation,
\begin{equation}
P^{(n+1)} = F(P^{(n)}),
\end{equation}
then the whole $T$ is given by a result after infinite iterations
with the proper initial condition $P^{(0)}$.
In this article, we set up this type of iterative method to calculate
the spontaneous mass generation.
As for the simplest toy example, see the Appendix A where the sum of 
geometric series is treated in this line of thought, which gives a 
clear view of our strategy. 

Using the iterative method we can evaluate directly the spontaneously
generated mass.
This is absolutely non-trivial, since the spontaneous mass generation
is nothing but the spontaneous symmetry breakdown of the chiral
symmetry and it is an issue of phase transition.
Phase transition of system is characterized by appearance of 
singularity and normally we have to make bypasses or deep consideration
to evaluate physical quantities related to the phase transition\cite{SSB}.

The only escape is that if we can regularize the system {\sl a la}
finite volume (finite number of degrees of freedom), then there is 
no singularity, nothing unphysical,  in any stage of regularization. 
We take the infinite volume
limit at the last of calculation, which finally generates singularities
in physical quantities and through such singularities, 
we obtain physically correct results, like the spontaneously generated mass.

Actually, our method described in this article can be regarded 
as a sort of this type of regularization realizing
singularity free direct calculation. The part $P^{(n)}$ is a regularized
quantity where $n$ represents the regularization parameter.
Moreover our $P^{(n)}$ has a good physical feature that it
assures the positivity of fluctuation at any $n$, which is easily 
lost in other approximation methods treating the spontaneous symmetry breakdown.

In Section 2, we set up the iterative summing up method of the all relevant
diagrams to give the dynamical mass in the Nambu--Jona-Lasinio model.
In Section 3, we show the spontaneous mass generation mechanism 
in this method and the results give the correct physical mass.
In Section 4, we apply the iterative method to calculate various physical 
functions. As for the Legendre effective potential, we obtain the
convex function automatically.  
In Section 5, we extend out method to the finite density system, where
the 1st order phase transition is expected to occur.
We find that our iteration fails to select the proper physical mass for
some special regions of parameters.
In Section 6, we treat the gauge theory and construct the similar
iterative method to sum up all ladder type diagrams, which works 
well to give the spontaneously generated mass.

\section{Iteration Method for NJL Model}

In this section, we adopt the Nambu--Jona-Lasinio (NJL) model\cite{nambu} 
and give a new
iterative method that directly sums up an infinite number of diagrams of the
standard perturbation theory in the bubble tree ($1/N$ leading)
 approximation. Using this method, we
demonstrate that the physically correct result is automatically obtained with
the precise critical coupling constant. 

The NJL model has four-fermi 
interactions among the massless fermions respecting the chiral invariance. We
add the bare mass $m_0\bar\psi\psi$ (explicitly breaking the chiral symmetry)
to the Lagrangian to make the standard perturbation
theory work properly,
\begin{equation}
 {\mathcal{L}}_{\rm{NJL}} = \bar{\psi}i\Slash{\partial}\psi+
 \frac{G}{2N} \left[\left({\bar{\psi}\psi}\right)^2 + 
 \left(\bar{\psi}i\gamma_{5}\psi\right)^2\right] - m_0\bar{\psi}\psi,
\end{equation}
where $N$ is the number of fermion flavors. 
This model is not renormalizable and we set the ultraviolet cutoff $\Lambda$.
The critical coupling constant for the spontaneous mass generation is
$G=4\pi^2$ and we also use the rescaled coupling constant 
$g=G/(4\pi^2)$.

\begin{figure}[!h]
\hbox to \hsize{\hfill\vbox{\hbox{{\Large$\displaystyle\sum_{\rm all}$}}\vskip-2mm}\hskip3mm
\includegraphics[width=40mm,clip]{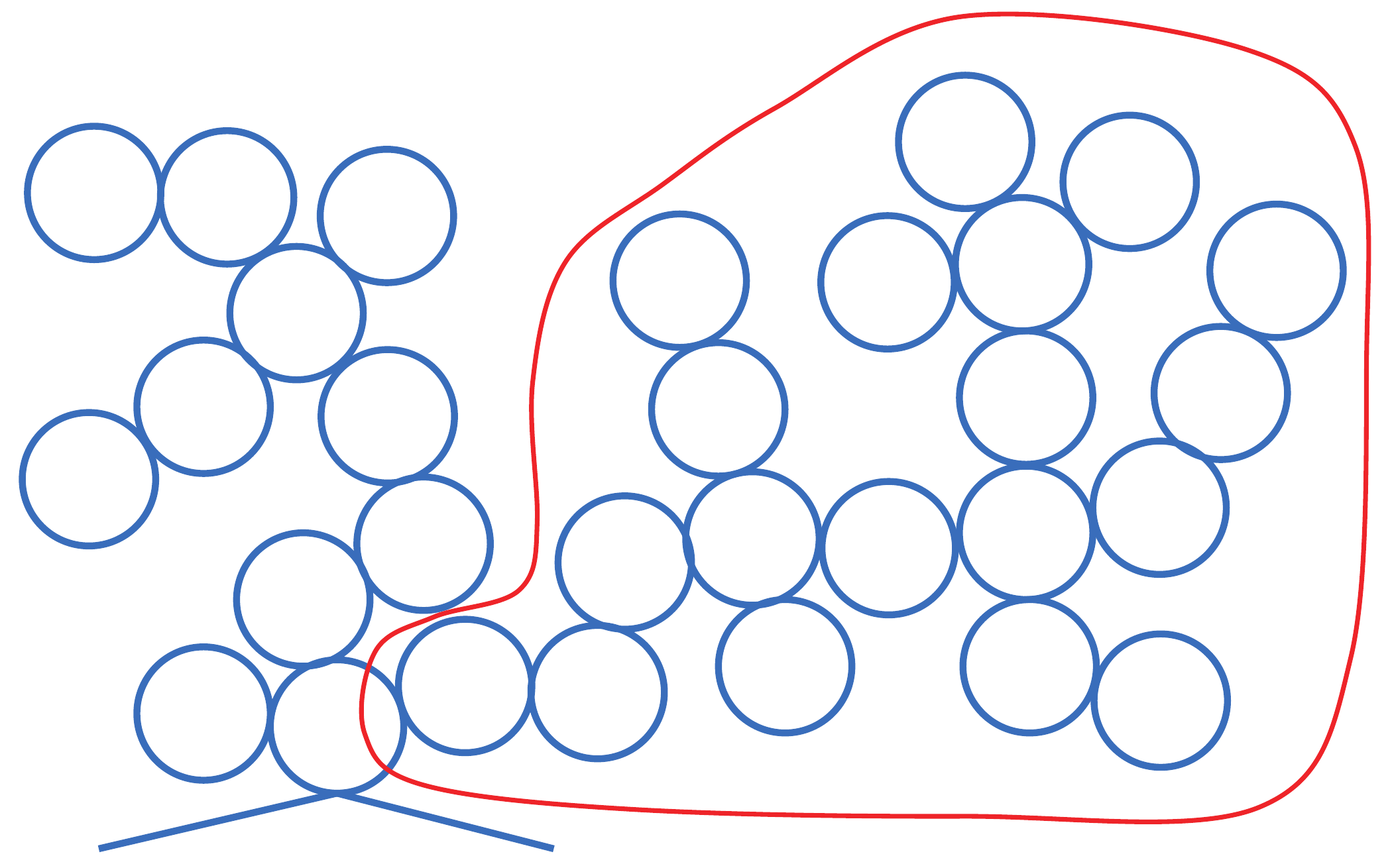}\hfill}
 \caption{Tree diagram with the whole as a part.}
 \label{fig:NJLT}
\end{figure}

We consider $1/N$ leading
contribution to the mass. Diagrammatically it is a sum of infinite diagrams
called {\sl tree} (Fig.\ref{fig:NJLT}), where considering the fermion-antifermion pair 
as a single meson, the tree diagrams are defined by those without any 
meson loops, or in other words we regard a series of loops as a {\sl fat}
propagator. Usual method is to set up a self-consistent equation satisfied by this infinite 
sum of diagrams. 

Above the critical coupling constant, there actually exists a
non-trivial solution which does not vanish after zero bare mass limit. However,
it is unclear that how the finite mass should come out of the infinite sum of
the diagrams while each diagram certainly vanishes at the zero bare mass limit.
Now, we set up a method to directly sum up the infinite number of diagrams and
show how the finite mass come out without any ambiguity nor singularity.

In the {\sl tree} type diagrams drawn in Fig.\ref{fig:NJLT}, 
we can easily find the whole sum resides as a part of the set.
The part surrounded by a line is equivalent to the whole sum.
This observation leads to the well-known self-consistent equation
given by the original Nambu--Jona-Lasinio paper.

\begin{figure}[h!]
 \CFF{90}{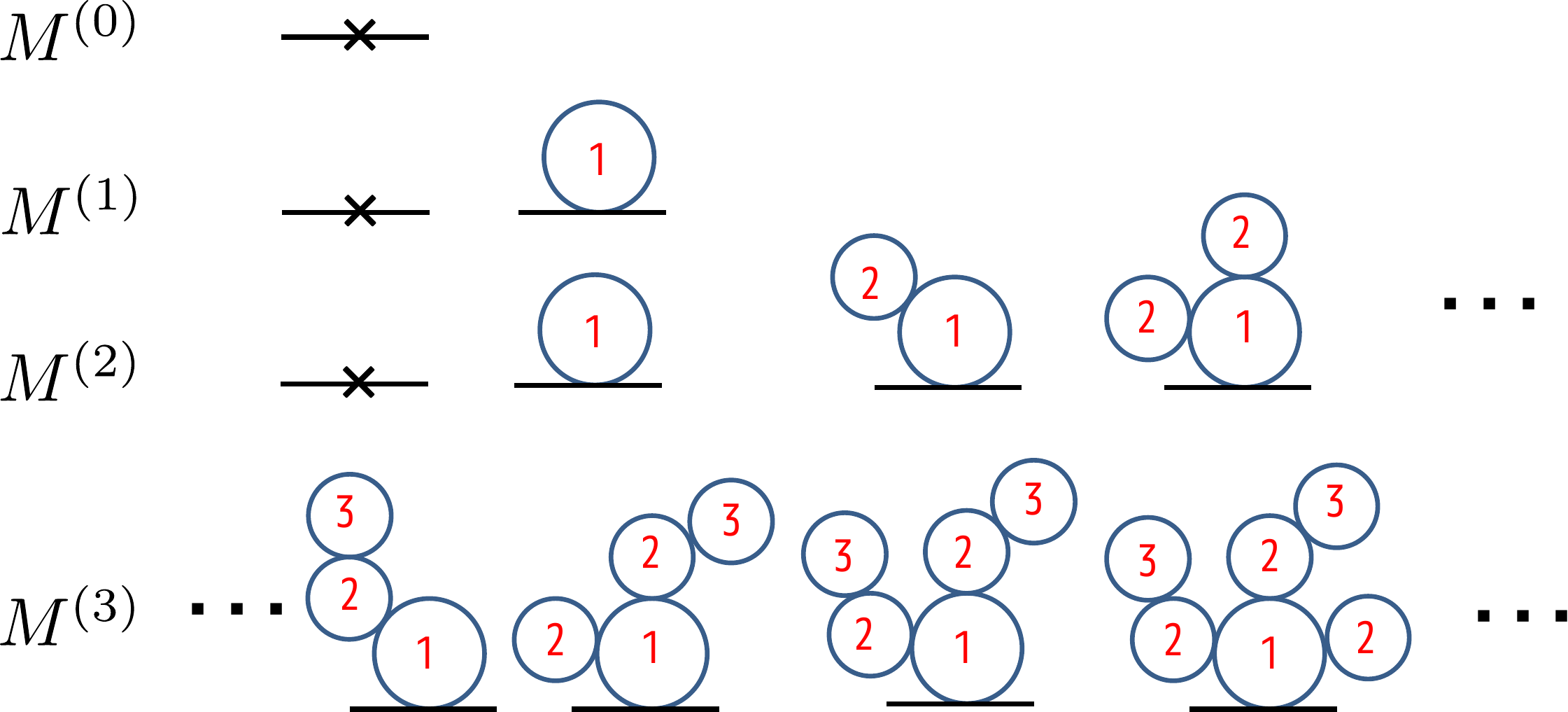}
 \caption{\label{fig:defNL} Definition of node length.}
\end{figure}

Now we set up the iterative method to sum up all {\sl tree} diagrams.
First of all, we classify diagrams in the {\sl tree} according to the node length of each
diagram. Node length of a diagram is defined by the maximum number of loops in a
continuous route from the mass external line towards the edge loop, or maximum number of nodes of {\sl fat}
propagator legs in the diagram. Fig.\ref{fig:defNL} shows the counting rule of node length
and classification of diagrams. 

\begin{figure}[h!]
 \CFF{70}{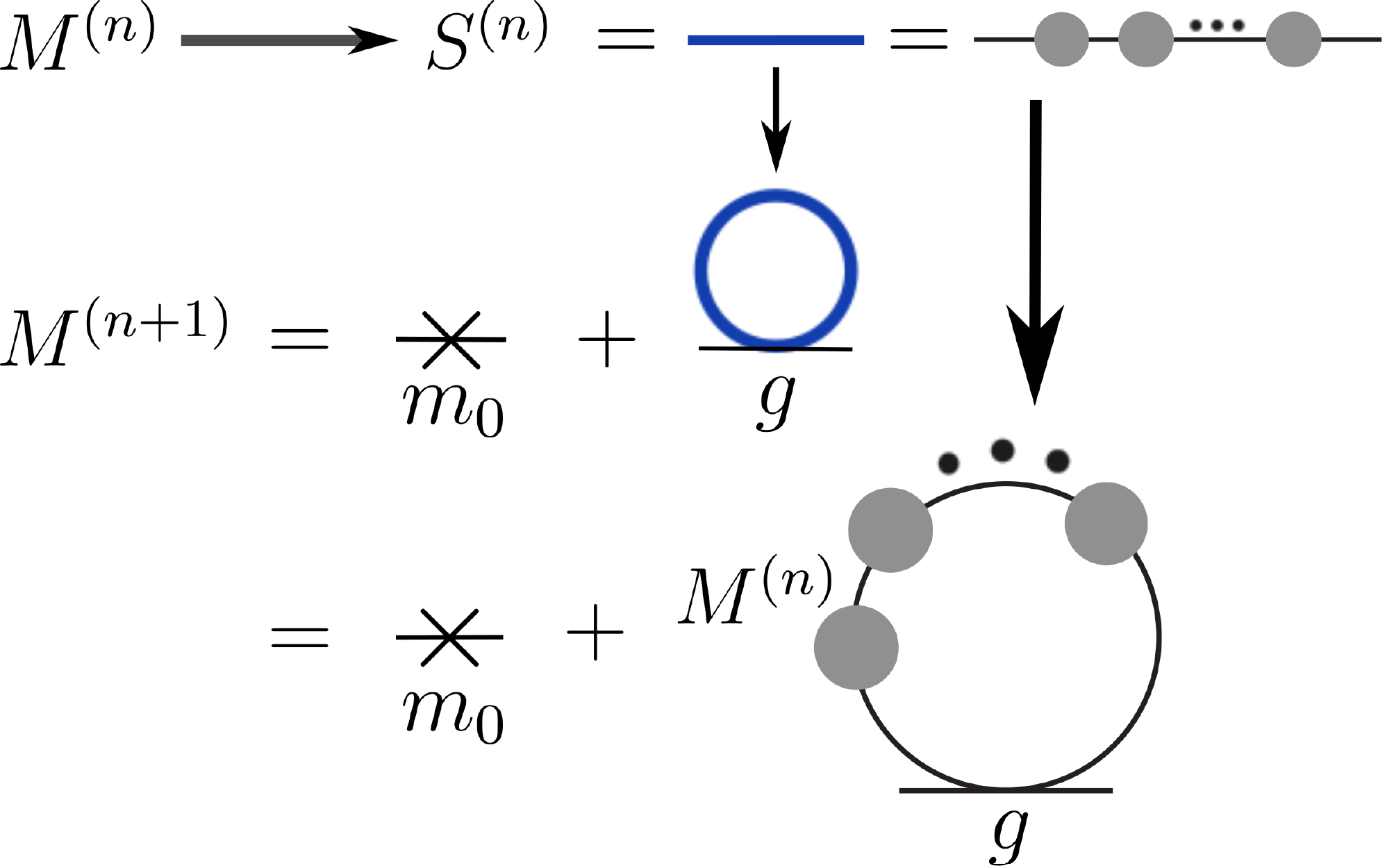}
 \caption{\label{fig:nodelengthiteration} Node length iteration.}
 \label{fig:NLIdef}
\end{figure}

Here we define $M^{(n)}$ as a sum of diagrams
whose node length is no greater than $n$. 
Then we find the iteration transformation 
to evaluate $M^{(n+1)}$ by using $M^{(n)}$, which is 
shown in Fig.\ref{fig:NLIdef}. 
Note that the qualitative structure of this iteration transformation
resembles much to that defined for the sum of geometric series (Appendix A).
At every iteration, we make $n+1$ node diagram using 
$n$ node diagram by looping the corresponding propagator and 
finally add the $0$-node term $m_0$ which is not contained
in the loop diagram.

The transformation function $F$,
\begin{equation}
 M^{(n+1)}=F\left(M^{(n)}\right),
\end{equation}
is given by the one loop integral as follows:
\begin{equation}
\begin{split}
F(M)&=m_0+4\pi^2 g \int \frac{d^4 p}{(2\pi)^4} {\rm tr}\left[  \frac{i}{\Slash{p}-M} \right]\\
&\hspace{-20pt}\scalebox{0.80}{$\xrightarrow{ \text{Wick rotation}} $}~m_0+16\pi^2 gM \int \frac{d^4 p_{\rm E}}{(2\pi)^4} \frac{1}{{p_{\rm E}}^2 + M^2} \\
&= m_0+16\pi^2 gM \int d \Omega \int_{0}^{\Lambda } \frac{d |p_{\rm E}|}{(2\pi)^4} \frac{|p_{\rm E}|^3}{|p_{\rm E}|^2 +M^2} \\
&=m_0+gM\left(1-M^2\log\left(1+M^{-2}\right)\right).
 \end{split}
\label{eq:NJLIT}
\end{equation}
Here all the variables are rescaled to be dimensionless taking the ultraviolet cutoff
$\Lambda$ as the mass unit. 

In conclusion, the total sum of the tree diagrams is obtained by $M^{(\infty)}$, that is, through infinitely many times of transformation by the same $F$.

\section{Mass Generation}
\begin{figure}[h!]
 \CFF{80}{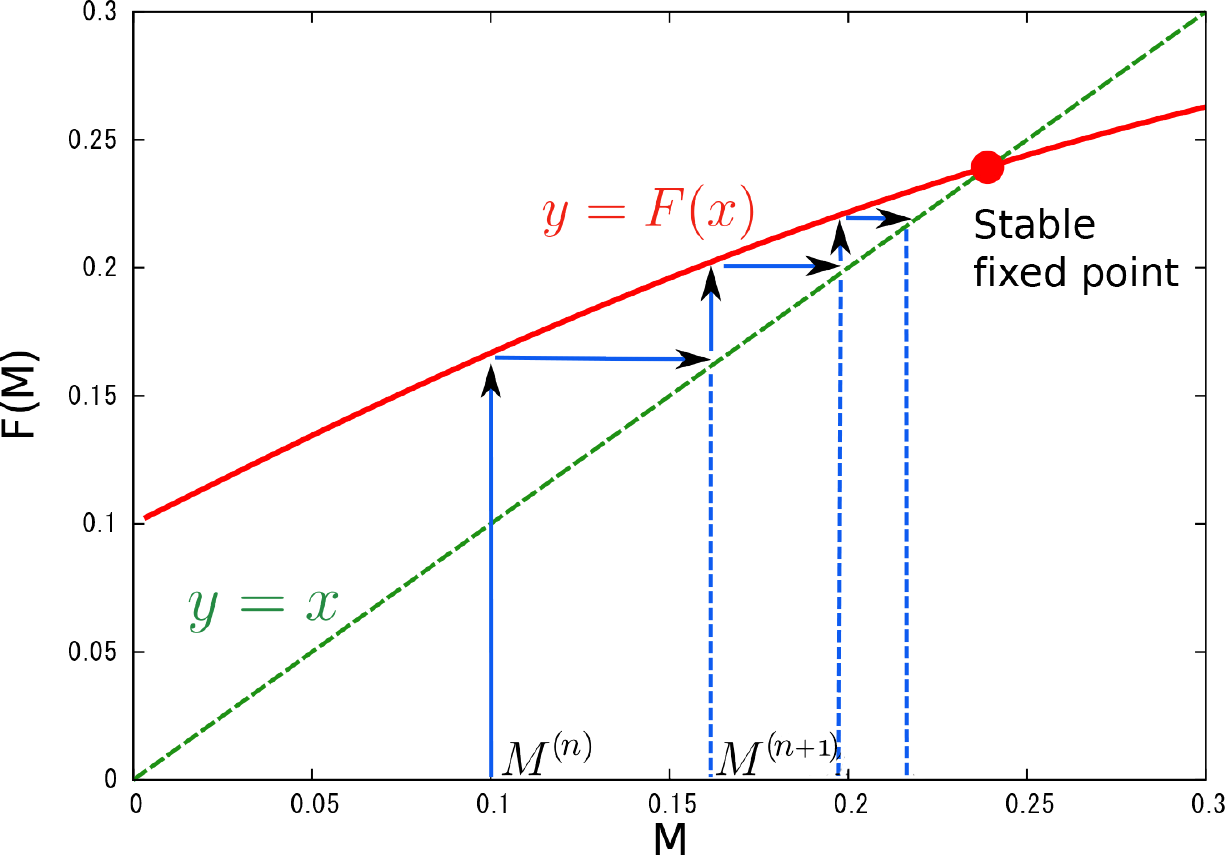}
\vskip5mm
 \CFF{80}{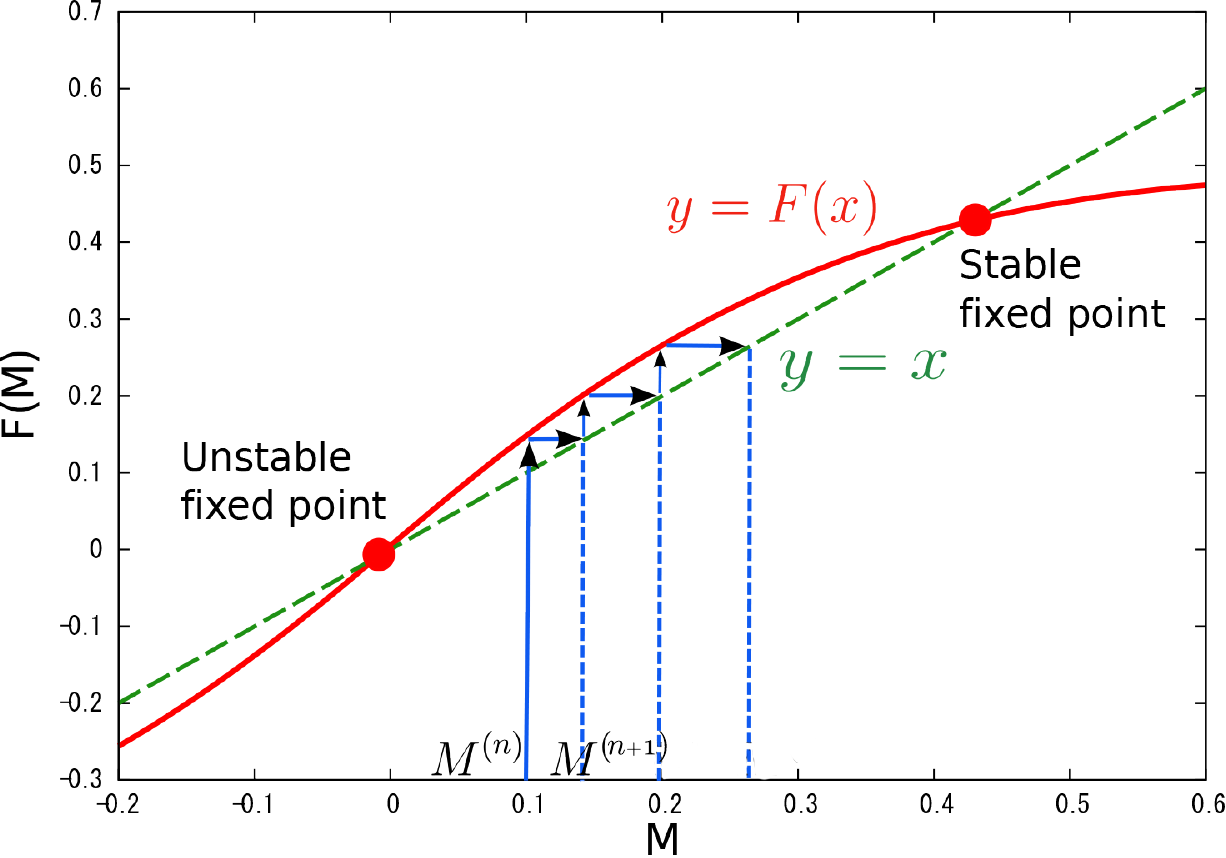}
 \caption{Iteration procedure for $g=0.7\text{(upper)}, 1.5\text{(lower).}$}
 \label{fig:NJLI}
\end{figure}

\begin{figure}[htbp]
 \centering
 \includegraphics[width=12cm]{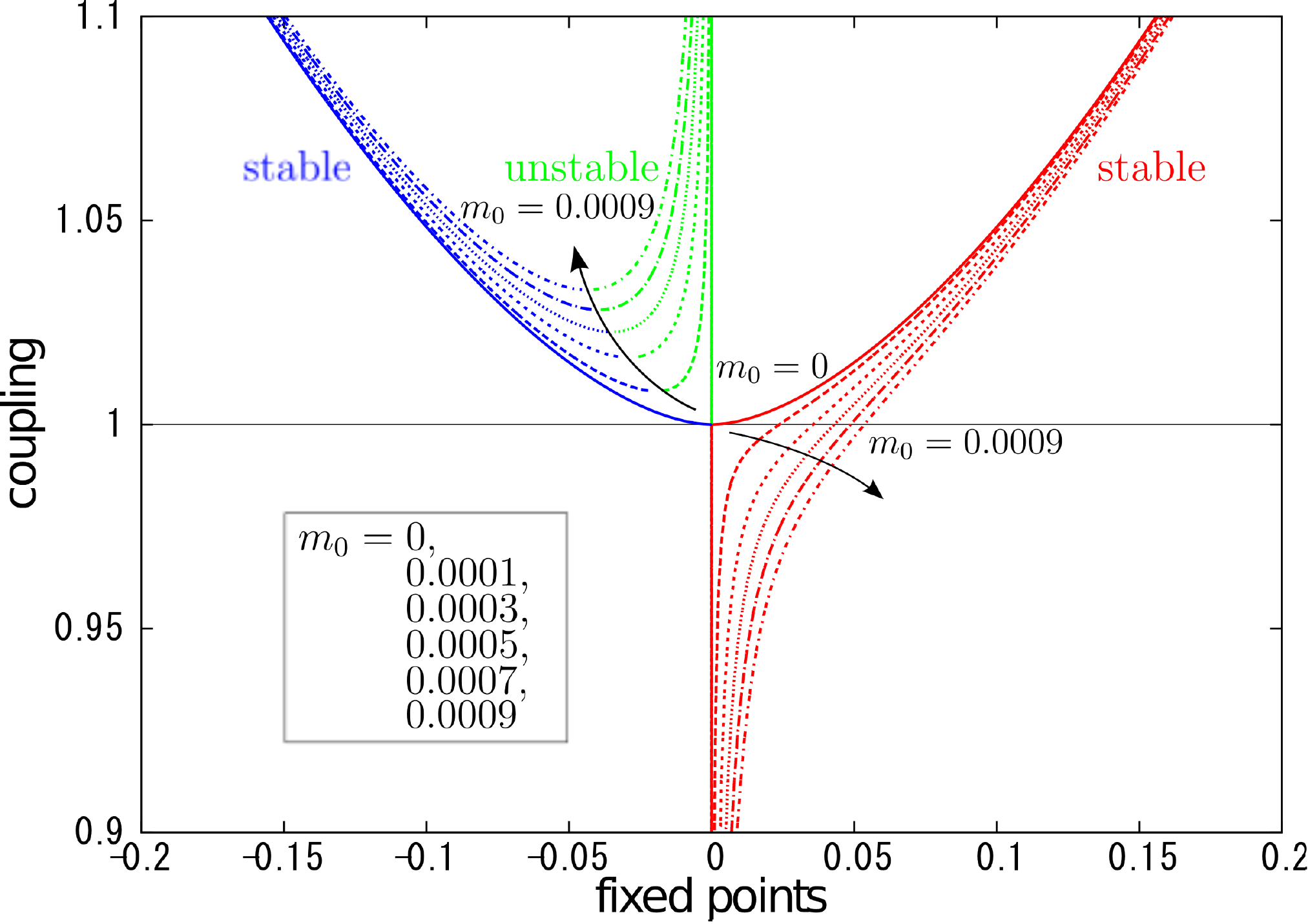}
 \caption{Position of fixed points vs $g, m_0$.}
 \label{fig:NJLFP}
\end{figure}

Iterative transformation here is best understood by a graphical method where the
transformation function $y=F(x)$ and a straight line $y=x$ are drawn as shown
in Fig.\ref{fig:NJLI}. Each iteration process can be drawn on this figure by a successive moves of point. In any case the iterative transformation finally reaches a
stable fixed point. Fixed points are crossing points between $y=F(x)$ and $y=x$,
and position of fixed points are shown in Fig.\ref{fig:NJLFP}.

In the weak coupling region ($g=0.7$) shown in the upper diagram in Fig.\ref{fig:NJLI}, 
there is only one fixed point near the origin and it is stable.
The iteration should start with the initial condition $M^{(0)}=m_0$ and 
it approaches to the fixed point.

When the coupling constant becomes strong, there occurs pair
creation of fixed points, one is stable and the other is unstable, 
which is seen in Fig.\ref{fig:NJLFP} where move of fixed point positions are
plotted for various $m_0$.

Then there
are two stable fixed points each of which has its attractive region, {\sl territory}. 
We must be careful about the initial starting point of iteration,
$m_0$, that is, the essential question is in which territory does it start.

In all figures, we use positive $m_0$, 
then the initial point exists in the
territory of the right-hand side stable fixed point, as seen in lower diagram in 
Fig.\ref{fig:NJLI}.
To prove this we investigate neighborhood of the origin, where
the transformation function takes the following form,
\begin{equation}
F(x)\simeq m_0 + kx, \ \ k>1.
\end{equation}
Then the fixed point near the origin $x^\star_0$ is obtained as
\begin{equation}
x^\star\simeq m_0 + k x^\star 
\longrightarrow x^\star\simeq \frac{m_0}{1-k},
\end{equation}
that is, $x^\star_0$ is negative. 
Therefore, for all region of the coupling constant, the physical result is
controlled by the right most stable fixed point in Fig.\ref{fig:NJLFP}. 

Note that the critical coupling constant is defined for $m_0=0$.
The criticality corresponds to the case that 
the gradient of iteration transformation function $F$ at the origin 
equals to unity. When it is larger than unity, there appears three fixed points
and one at the origin becomes unstable.
The gradient is quickly evaluated as 
\begin{equation}
F^\prime(0)= g,
\end{equation}
and therefore the critical coupling constant is obtained  as
\begin{equation}
g_{\rm c}=1.
\end{equation}

\begin{figure}[h!]
 \CFF{80}{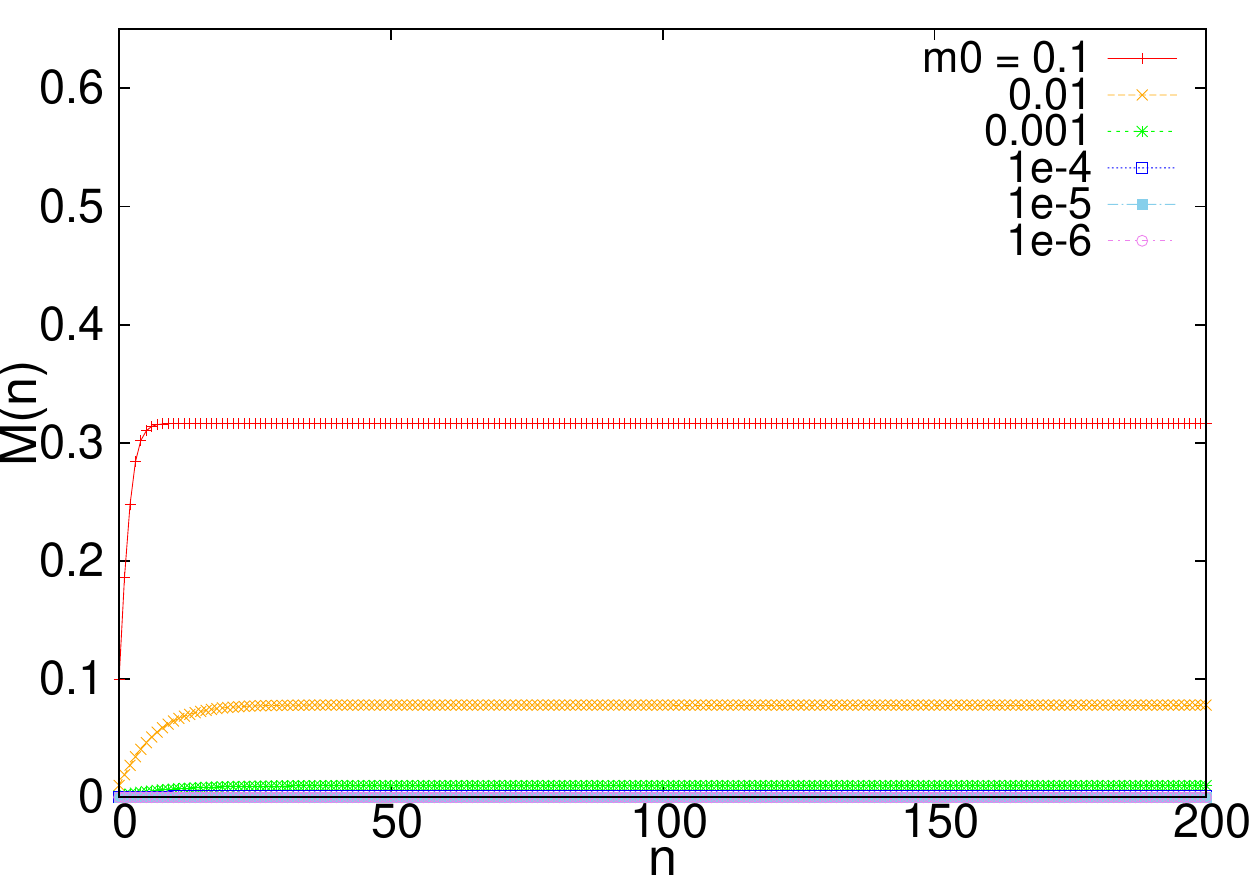}
 \CFF{80}{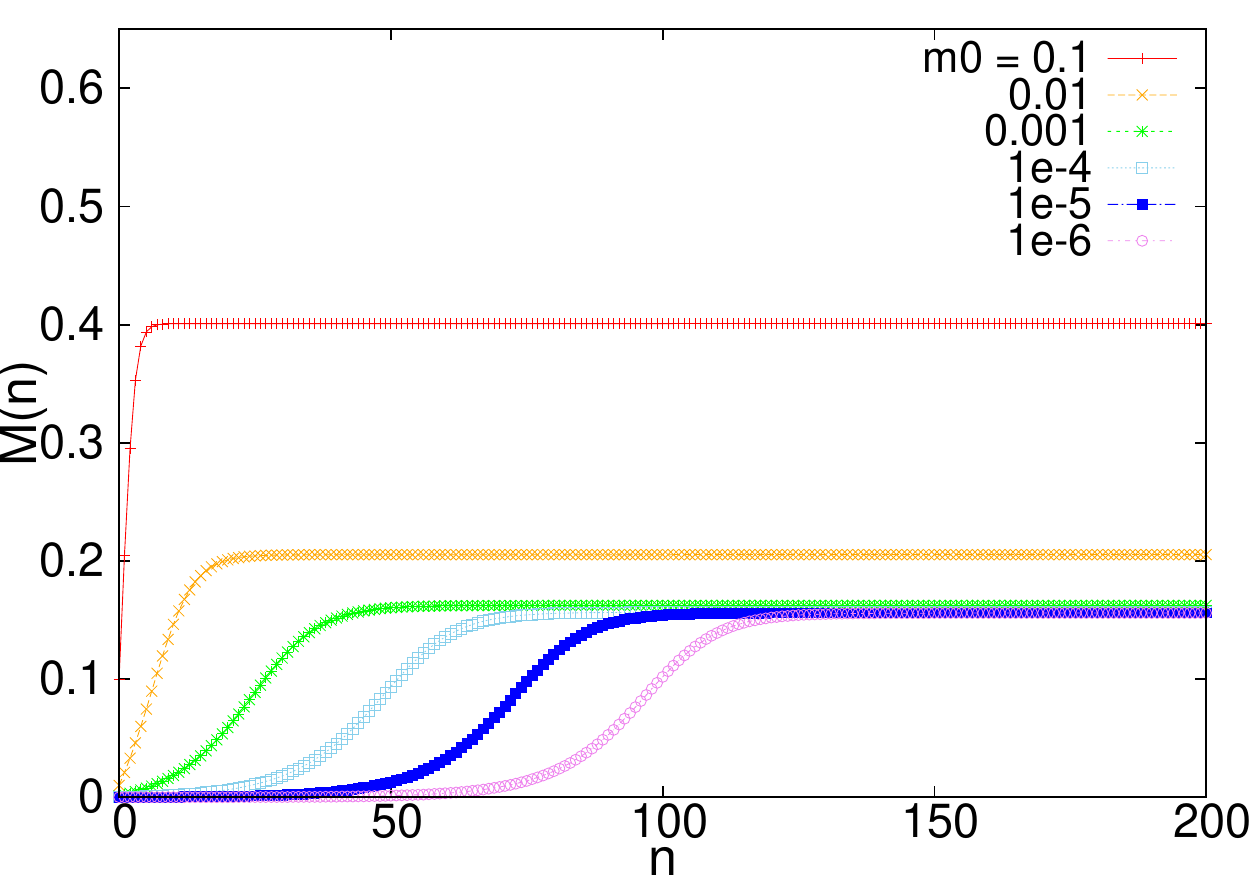}
 \caption{Mass generation procedure for $g=0.9\text{(upper)}, 1.1\text{(lower)}$.}
 \label{fig:NJLM}
\end{figure}

Let's see some features of mass generation with respect to the node length $n$ in
Fig.\ref{fig:NJLM}. In the weak coupling case (upper diagram), 
the dynamical mass is generated rather
quickly at low $n$ and becomes constant, which should be called the perturbative
characteristics. By decreasing the bare mass the final mass goes to zero. 

In the
strong coupling case (lower diagram), the generation of the dynamical mass depends strongly
on the bare mass, and it is mainly generated at some narrow range of node 
length. Decreasing the bare mass, the region of mass generating node length
becomes large, but the output mass is almost constant, which means the
spontaneous mass generation. 

It is also seen that the shape of generation curves
look the same form, just displacement in the node length space. These features
are readily understandable by the iterative nature of our calculation well seen in
the lower diagram in Fig.\ref{fig:NJLI}.
The move of iterated points is characterized from the unstable fixed point to 
the stable fixed point. 
When taking smaller bare mass, the starting point is nearer to the unstable
fixed point and thus the growing up is delayed, thus larger node length region is important.
The quick growing region is the mid of the two fixed points, and this assures the
iteration behavior of mass growing is quite similar independent of $m_0$,
just the translation in the node length space. 

\begin{figure}
\CFF{85}{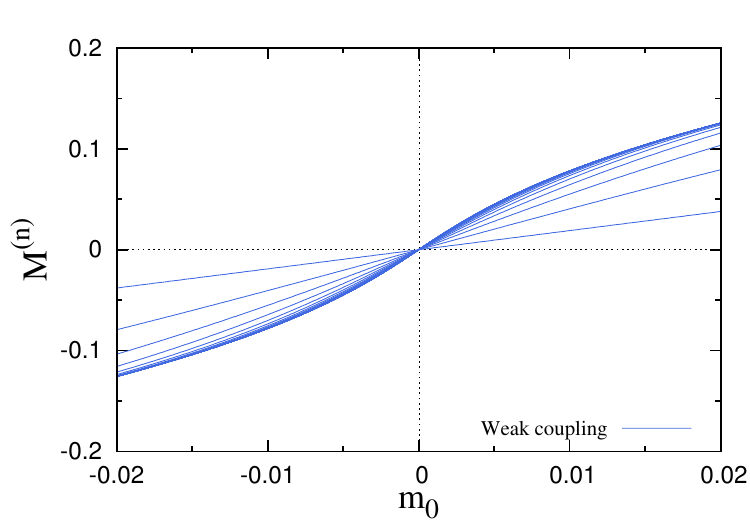}
\CFF{85}{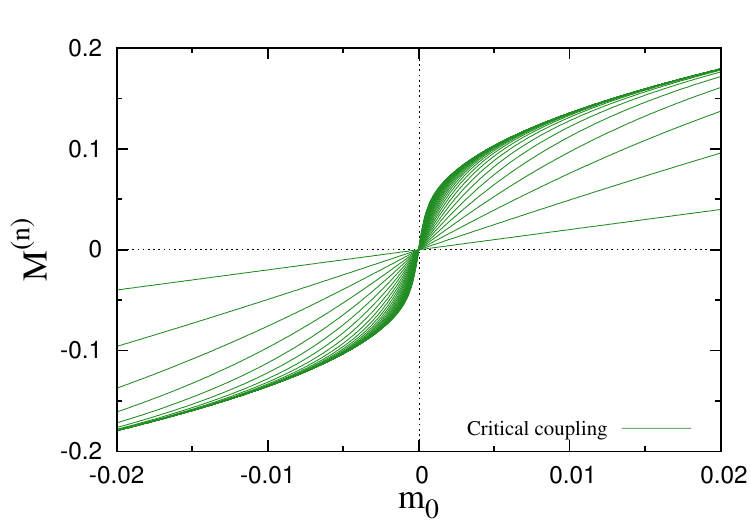}
\CFF{85}{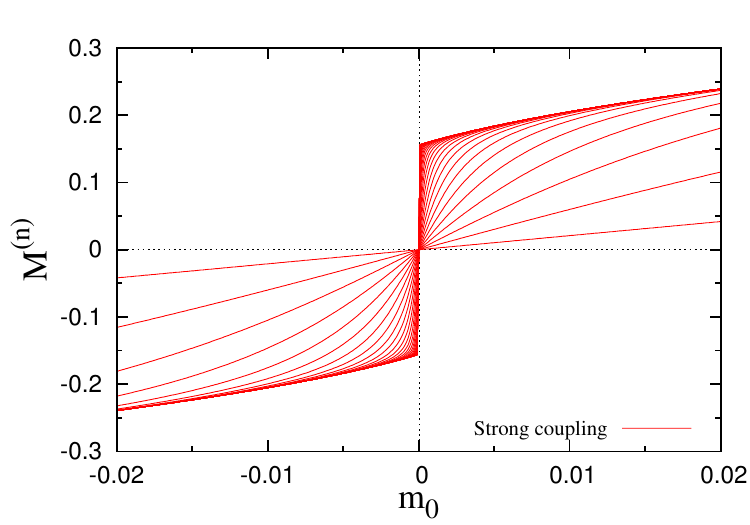}
\label{fig:nodelengthiteratedM}
\caption{Iterated $M^{(n)}$ for $g=0.9\text{(upper)}, 1.0\text{(mid)}, 1.1\text{(lower)}$.}
\label{fig:NJLMm}
\end{figure}

In Fig.\ref{fig:NJLMm}, we plot iterative development of mass as a function of $m_0$.
The upper diagram is below critical and vanishing bare mass limit of $M$ 
is always vanishing.

The mid diagram is just on critical, and we see the appearance of infinite slope
at the origin. The slope of $M$ with respect to $m_0$ is nothing but the
susceptibility of the chiral condensate. Therefore, the susceptibility becomes
divergent at $n\rightarrow\infty$, which is the characteristics of the 2nd order
phase transition.

The lower diagram is super critical. Here vanishing bare mass limit of $M$
is vanishing also, for any finite $n$.
This implies that for any finite node length $n$, the spontaneous mass
generation does not occur. Of course, this is also well understandable
if we imagine the iteration procedure in Fig.\ref{fig:NJLI}.
However, if we change the order of limit, that is, keeping the non-vanishing 
bare mass, we take the infinite node length limit first as
\begin{equation}
\lim_{m_0\rightarrow 0} [\lim_{n\rightarrow\infty} M^{(n)}(m_0)],
\end{equation}
then it gives a non-vanishing value. This should be regarded as the
spontaneously generated mass.

\begin{figure}
\CFF{90}{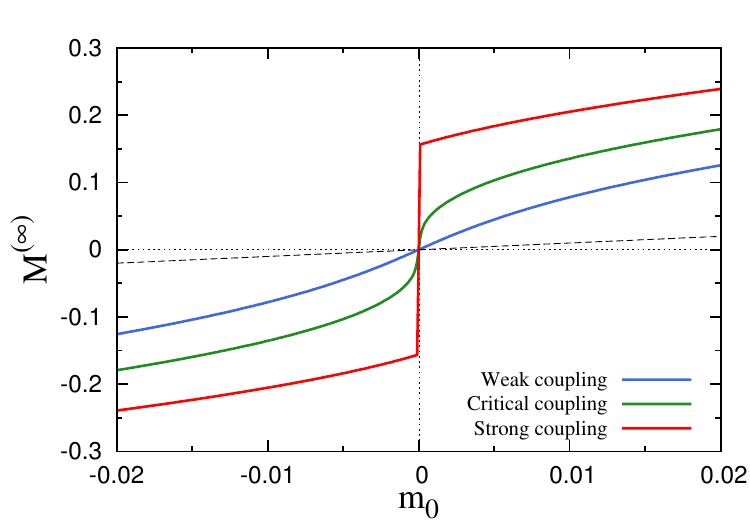}
\caption{$M^{(\infty)}$ as a function of $m_0$ for $g=0.9, 1.0, 1.1$.}
\label{fig:NJLMinfty}
\end{figure}

In this way, we conclude the dynamical mass given by the iterative method
as shown in Fig.\ref{fig:NJLMinfty}.

\begin{figure}[h!]
 \CFF{70}{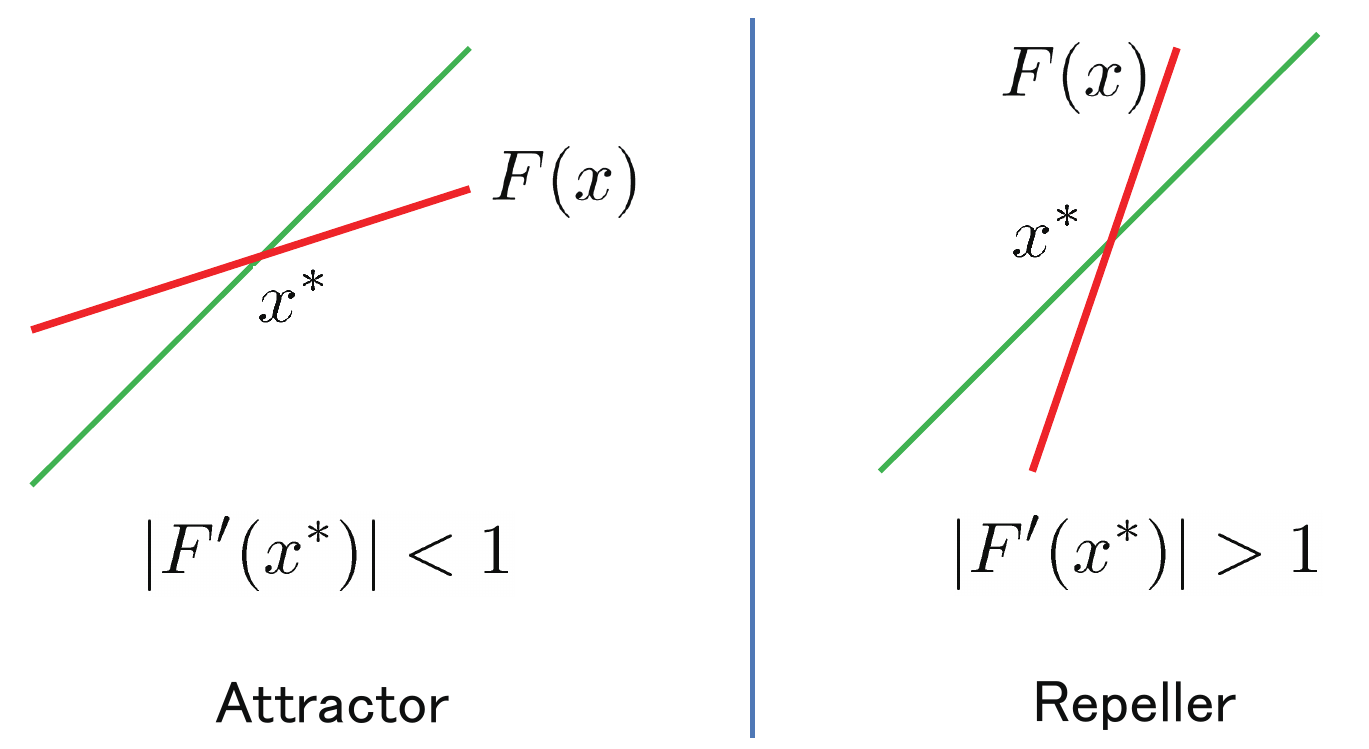}
 \caption{Eigenvalues of linearized transformation.}
 \label{fig:EVL}
\end{figure}

Here we comment on an implicit relation between unstableness of 
fixed point and the physicality condition.
We rewrite the iteration transformation as
\begin{align}
 M^{(n+1)}= F (M^{(n)})=m_0 +F_0 (M^{(n)})\ , 
\end{align}
then fixed point $M^\ast$ satisfies
\begin{align}
 M^{\ast}= F (M^{\ast})=m_0 +F_0 (M^{\ast})\ .
\end{align}
The slope of function $F$ at a fixed point gives the
eigenvalue of the transformation linearized around the fixed point and thus
it determines the stability of it as shown in Fig.\ref{fig:EVL}.

The slope is calculated as
\begin{align}
\left. \frac{dF(M^{\ast})}{dM^{\ast}} \right|_{m_0 \rm{fixed}}&=\frac{dF_0 (M^{\ast})}{dM^{\ast}}=1-\frac{d m_0 (M^{\ast})}{dM^{\ast}}.
\end{align}
Therefore, the unstable fixed point,
\begin{equation}
F^\prime(M^{\ast}) > 1,
\end{equation}
corresponds to the negative derivative,
\begin{equation}
\frac{d m_0 (M^{\ast})}{dM^{\ast}} <0.
\end{equation}
On the other hand in our model of NJL ladder, inverse of this derivative
corresponds to the susceptibility $\chi$ of $\bar\psi\psi$ as
\begin{equation} 
\frac{dM^\ast} {d m_0} = G\chi +1, 
\end{equation}
which is derived by Eq.(\ref{eq:phiMm}).
Then the unstable fixed point corresponds to the negative $\chi$, that is, 
negative fluctuation of $\bar\psi\psi$, which means the instability of the
vacuum and absolutely unphysical solution. Inversely, the normal vacuum 
($\chi>0$) assures the stability of fixed point $|F^\prime(x^\ast)|<1$.

\section{Free Energy and Effective Potential}

In this section, we calculate the free energy and the effective potential
using the node length iteration. We define the free energy through the
logarithm of the partition function as a function of the bare mass $m_0$:
\begin{equation}
 W(m_0) \equiv \ln Z(m_0),
\end{equation}
where the partition function is given by
\begin{equation}
 Z(m_0)=\int {\mathcal{D}}\psi{\mathcal{D}}\bar\psi
  \exp \left(-\int d^4x_{\rm E}{\mathcal{L}}^{\rm E}_{\rm NJL}
\right).
\end{equation}
The free energy $W(m_0)$ is the generating function of the connected
Green function. Particularly its first derivative represents the vacuum
expectation value of operator $\bar\psi \psi$ as follows:
\begin{eqnarray}
 \frac{\partial W(m_0)}{\partial m_0}
  &=& \left<\int d^4x_{\rm E}\bar\psi(x_{\rm E})\psi(x_{\rm E}) \right>_{m_0} = \int d^4x_{\rm E}\left<\bar\psi(0)\psi(0)\right>_{m_0} \\
  &=& \Omega\left<\bar\psi \psi\right>_{m_0} \equiv \Phi, \nonumber
\end{eqnarray}
where $\left<\cdots\right>_{m_0}$ denotes the vacuum expectation value
and due to the translational invariance of the vacuum there is no
$x_{\rm E}$ dependence of $\left<\bar\psi \psi\right>_{m_0}$.

We introduce the Legendre transform of $W(m_0)$ by defining $\Gamma(\Phi)$,
\begin{equation}
 \Gamma(\Phi) \equiv -W(m_0)+m_0\Phi\ .
\end{equation}
Its derivative gives the bare mass in turn,
\begin{equation}
 \frac{\partial \Gamma(\Phi)}{\partial \Phi} =
  - \frac{\partial m_0}{\partial \Phi}\frac{\partial W(m_0)}{\partial m_0}
  + \frac{\partial m_0}{\partial \Phi}\Phi + m_0 = m_0.
\end{equation}
We move to the density of all these variables as follows:
\begin{equation}
 w(m_0) \equiv \frac{W(m_0)}{N\Omega}\ , \ \ 
 \phi\equiv \frac{\Phi}{N\Omega}\ ,\ \ 
 V_{\rm L}(\Phi) \equiv \frac{\Gamma(\Phi)}{N\Omega}\ ,\ \ 
 \frac{\partial V_{\rm L}(\phi)}{\partial \phi} = m_0.
\end{equation}
We also define the dimensionless variable $\tilde\phi$ by 
\begin{eqnarray}
 \tilde\phi &\equiv& \frac{\phi}{\Lambda^3} = \frac{\Psi}{\Omega N \Lambda^3} \\
  &=& \frac{1}{4\pi^2}{\tilde M}\left[1-{\tilde M}^2\ln\{1+{\tilde M}^{-2}\}\right].
\end{eqnarray}
Hereafter we omit the tilde mark for the dimensionless variables and 
find the simple relation, $\phi$ as a function of $M$:
\begin{equation}
 \phi= \frac{M-m_0}{G}.
 \label{eq:phiMm}
\end{equation}

\begin{figure}
\CFF{85}{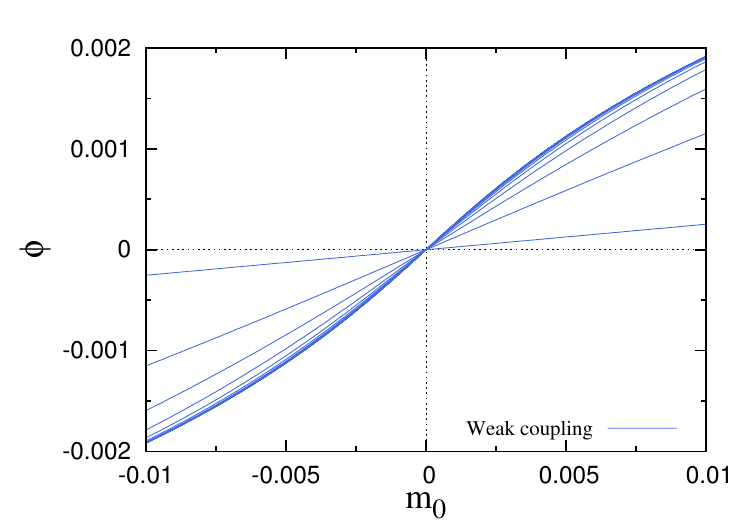}
\CFF{85}{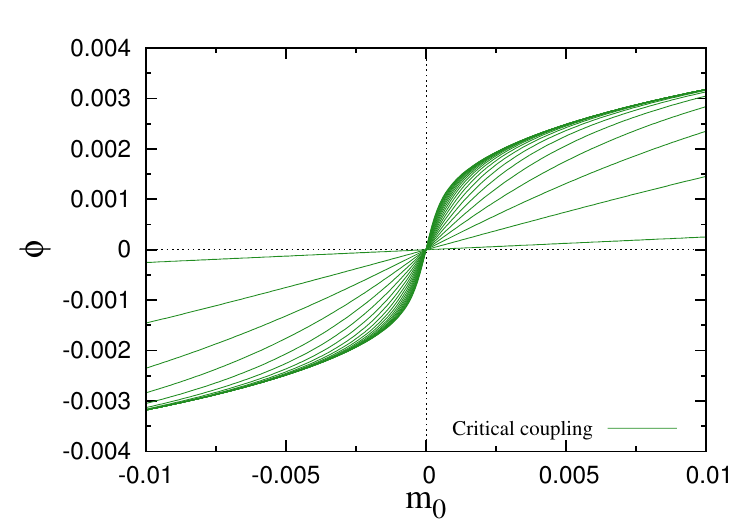}
\CFF{85}{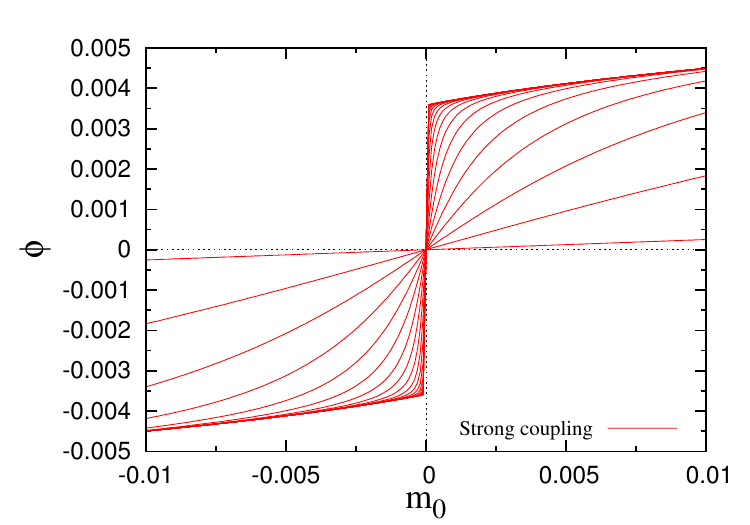}
\caption{$\phi^{(n)}$ as a function of $m_0$ for $g=0.9\text{(upper)}, 1.0\text{(mid)}, 1.1\text{(lower)}$.}
\label{fig:NJLphi}
\end{figure}

\begin{figure}
\CFF{87}{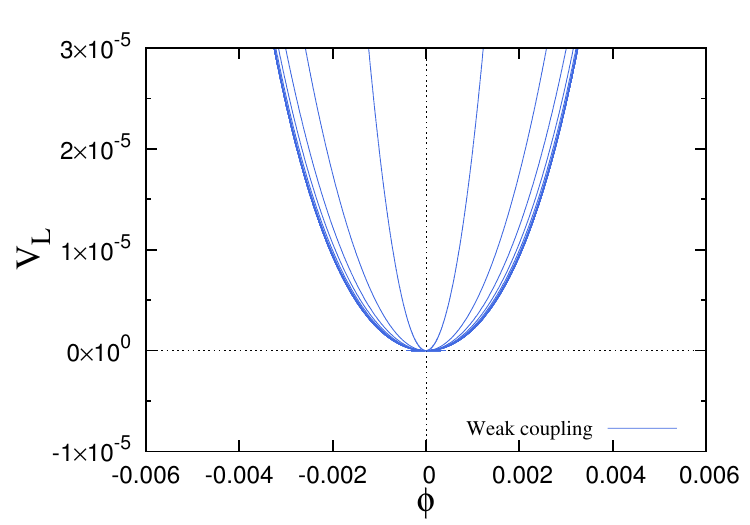}
\CFF{87}{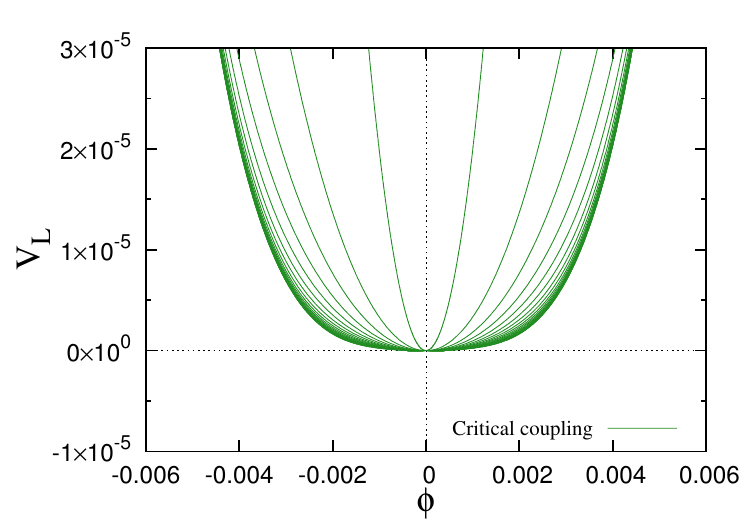}
\CFF{87}{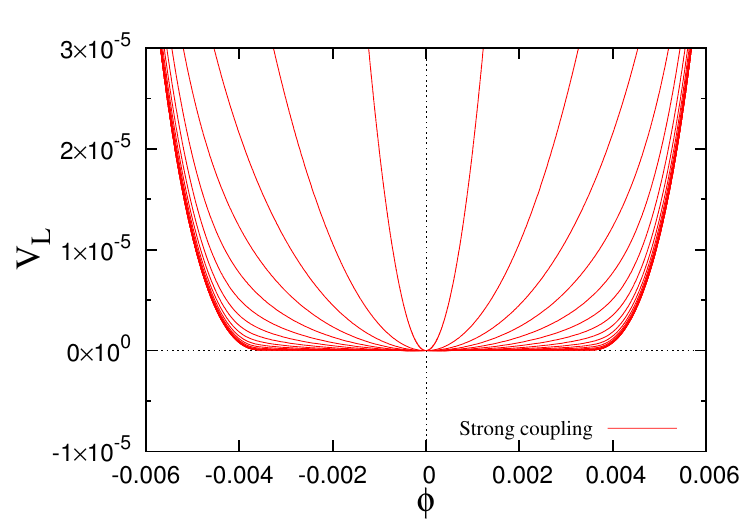}
\caption{Effective potential $V^{(n)}_{\rm L}$ for $g=0.9\text{(upper)}, 1.0\text{(mid)}, 1.1\text{(lower)}$.}
\label{fig:NJLVL}
\end{figure}

Now we apply the node length iteration method to this relation and
define node length iterated $\phi$, $w$ and $V_{\rm L}$ as follows:
\begin{equation}
 \phi^{(n)}(m_0)\equiv\frac{M^{(n)}-m_0}{G}\ ,\ \ 
\end{equation}
\begin{equation}
\frac{\partial w^{(n)}(m_0)}{\partial m_0}= \phi^{(n)}(m_0)\ ,\ \ 
V^{(n)}_{\rm L}(\phi^{(n)})=-w^{(n)}(m_0)+ m_0 \phi^{(n)}.
\end{equation}
Using these relations, we first calculate $\phi^{(n)}$ as a function of $m_0$,
and then integrate it to have the function $w^{(n)}(m_0)$. Finally we 
obtain the Legendre effective potential function $V^{(n)}_{\rm L}$.  

All these results are plotted in Fig.\ref{fig:NJLphi} and Fig.\ref{fig:NJLVL}.
Note that calculated Legendre effective potential are perfectly convex
at any $n$, and therefore also convex at $n\rightarrow\infty$.
To prove this property we recall the iteration transformation in 
Eq.(\ref{eq:NJLIT}), 
\begin{equation}
M^{(n+1)}= F(M^{(n)})\ , F(M)=m_0+gM(1-M^2\log(1+M^{-2})).
\end{equation}
Differentiate both sides of this transformation with respect to 
$m_0$, we have
\begin{equation}
\frac{\partial M^{(n+1)}}{ \partial m_0} 
= 1 + F^\prime( M^{(n)})
  \frac{\partial M^{(n)}}{\partial m_0}.
\end{equation}
The derivative $F^\prime$ is found to be positive for the normal
physical region $|M| \leq 0.7$.
Taking account of the initial condition
\begin{equation}
\frac{\partial M^{(0)}}{\partial m_0} =1,
\end{equation}
we get 
\begin{equation}
\frac{\partial M^{(n)}}{\partial m_0} > 1, 
\end{equation}
for any $n$.
This inequality assures that at any $n$, the fluctuation of 
$\phi$ is always positive and hence the convexity of the Legendre
effective potential.

These physically correct results are automatically obtained in our method of
iterative evaluation of all physical variables.
As is mentioned in the introduction, the iterative method here realizes 
something like the finite volume regularization.

\section{Finite Density System}

In this section, we explore the finite density system. We add the chemical
potential ($\mu$) term to the Nambu--Jona-Lasinio model Lagrangian as follows: 
\begin{equation}
   {\cal L_{\rm NJL}} = \bar{\psi}i{\Slash\partial}\psi + \mu{\bar\psi}\gamma^0\psi
   - m_0{\bar\psi}\psi + 
  \frac{G}{2N} \left[\left({\bar{\psi}\psi}\right)^2 +
  \left(\bar{\psi}i\gamma_{5}\psi\right)^2\right].
\end{equation}
Then the fermion propagator is changed as
\begin{eqnarray}
 \frac{i}{{\Slash p}-m_0} \rightarrow \frac{i}{{\Slash p}+\mu\gamma^0-m_0},
\end{eqnarray}
and the basic one-loop integral appearing in our iteration method 
now takes the following form:
\begin{equation}
 \Sigma =4iG \int \frac{d^4p}{(2\pi)^4}\frac{M}{p^2 + 2\mu p^0 + \mu^2 - M^2}\ .
\end{equation}

So far our model has an ultraviolet cutoff $\Lambda$ and we take the
four-dimensional isotropic cutoff. Hereafter we take the so-called
three-dimensional cutoff since it better fits to the case with
finite temperature. The energy (time) component has no cutoff and only the
momentum (space) components have the cutoff $\Lambda$, that is, the integration
range for the energy and momentum takes the following ranges respectively,
\begin{eqnarray}
 \left\{
  \begin{array}{l}
   p^0:-\infty \rightarrow \infty\ , \\
   |{\bm p}| :0 \rightarrow \Lambda\ .
  \end{array}
 \right.
\end{eqnarray}

Transforming to the Euclidean coordinate, we have
\begin{equation}
\Sigma
= \frac{4GM}{(2\pi)^4} \int_{0}^{\Lambda}d^3p \int_{-\infty}^{\infty}dp^4 \frac{1}{\{p^4-i(\mu+\omega_{\bm p})\}\{p^4-i(\mu-\omega_{\bm p})\}}\ ,
\end{equation}
where $\omega_{\bm p}\equiv\sqrt{{\bm p}^2 +M^2}$.
We integrate the time component by using the residue theorem to
have
\begin{equation}
\Sigma= \frac{GM}{\pi^2}\int_{0}^{\Lambda}dp\ \theta(\sqrt{p^2+M^2}-\mu)\frac{p^2}{\sqrt{p^2+M^2}}\ .
\end{equation}
Finally, we obtain the loop integral:
\begin{align}
 \Sigma
 &=  2gM\theta(\sqrt{1+M^2}-\mu)\left[
 \theta(|M|-\mu)\left\{
 \sqrt{1+M^2}+M^2\ln\frac{|M|}{1+\sqrt{1+M^2}}\right\}\right. \nonumber\\
 &{}\left.+\theta(\mu-|M|)\left\{\sqrt{1+M^2}-\mu\sqrt{\mu^2-M^2}
 +M^2\ln\frac{\mu+\sqrt{\mu^2-M^2}}{1+\sqrt{1+M^2}}\right\}
 \right],
\end{align}
where all variables are rescaled to be dimensionless with unit $\Lambda$
and we use the rescaled coupling constant $g=G/(2\pi^2)$.
Note that in our three-dimensional cutoff the critical coupling constant
(for $\mu=0$) is $2\pi^2$, just a half of that of the four-dimensional
cutoff. This change of the physical criticality comes from the fact that 
the NJL model is not a renormalizable theory 
and the cutoff scheme is a part of definition of the theory.

We set up the node length iteration using the above 
cutoffed integral $\Sigma$ as follows:
\begin{align}
 M^{(n+1)} = F(M^{(n)}) = m_0 + \Sigma(M^{(n)})
\end{align}

\begin{figure}[h]
\CFF{70}{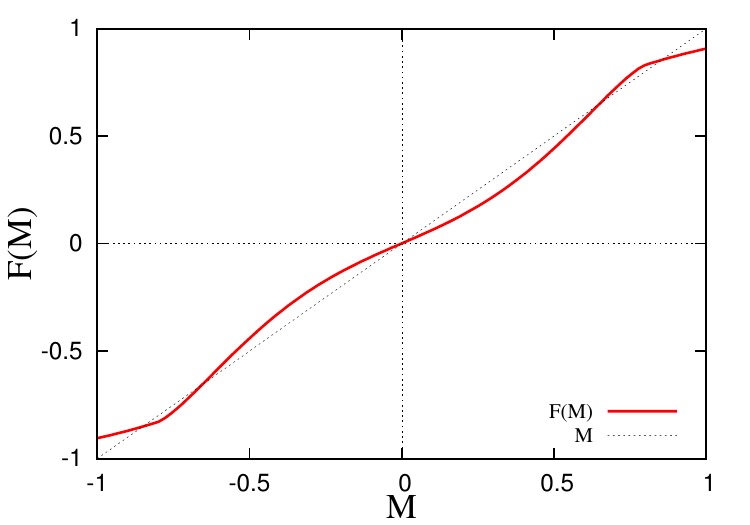}
\caption{Iteration function for $g=0.85, \mu=0.7, m_0=0$.}
\label{fig:ifmu}
\end{figure}

The iteration function $F(M)$ can have five fixed points at most in case of
strong coupling and low chemical potential.
We plot an example of the iteration function 
in Fig.\ref{fig:ifmu} where five crossing points are observed.

\begin{figure}[!h]
\CFF{90}{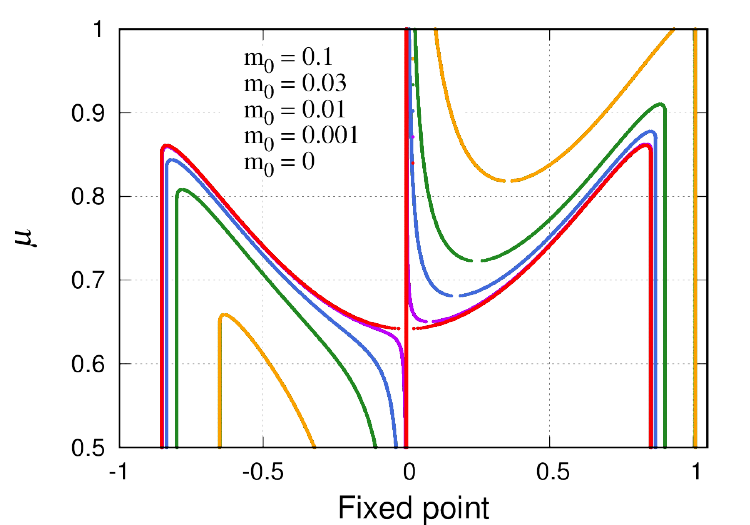}
\caption{Fixed point map for $g=0.85$.}
\label{fig:bifmu}
\end{figure}

\begin{figure}[!h]
\CFF{90}{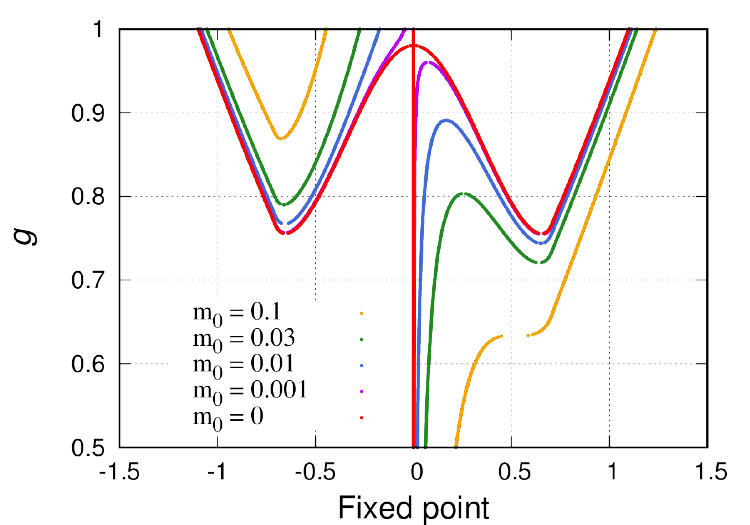}
\caption{Fixed point map for $\mu=0.7$.}
\label{fig:bifG}
\end{figure}

The total structure of the fixed point map is drawn in Fig.\ref{fig:bifmu} for
fixed $g=0.85$ and in Fig.\ref{fig:bifG} for fixed $\mu=0.7$.
These figures clearly show that we encounter the 1st order phase transition.
In Fig.\ref{fig:bifmu},  
looking down in the direction of the chemical potential from 1, 
or in Fig.\ref{fig:bifG}, 
looking up in the direction of $g$ from 0, at some point 
there appears pair creation of fixed points far from the origin, 
one is stable and the other is unstable, while 
the origin still survives as a stable fixed point.
This region with five fixed points correspond to the 
triple-well image of the potential.

\begin{figure}
\centerline{
\FF{65}{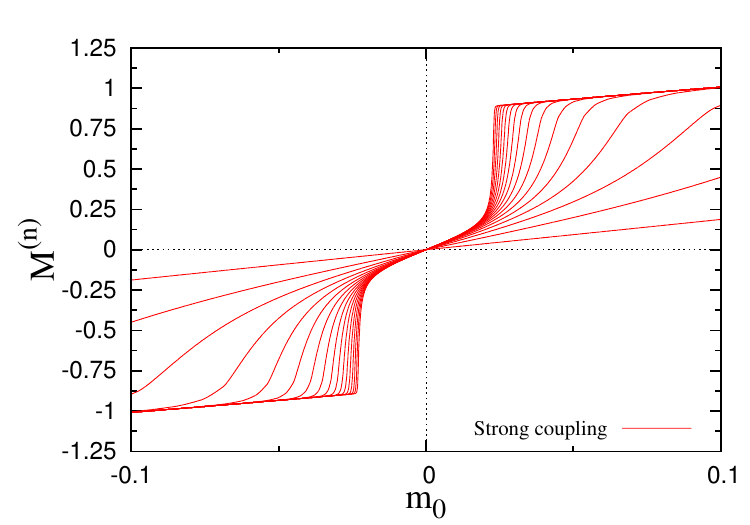}
\FF{65}{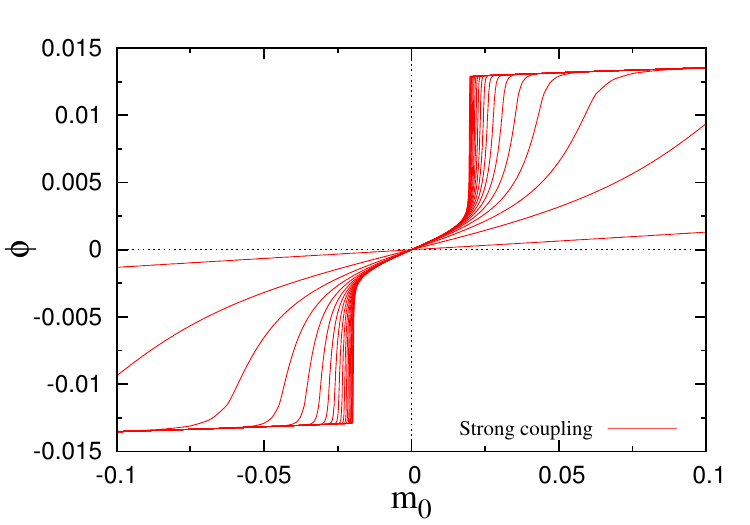}
}
\caption{Iteration of $M^{(n)}, \phi^{(n)}$ 
as a function of $m_0$ for $g=0.85, \mu=0.7$.}
\label{fig:Mmuphi}
\end{figure}

\begin{figure}[!h]
\CFF{80}{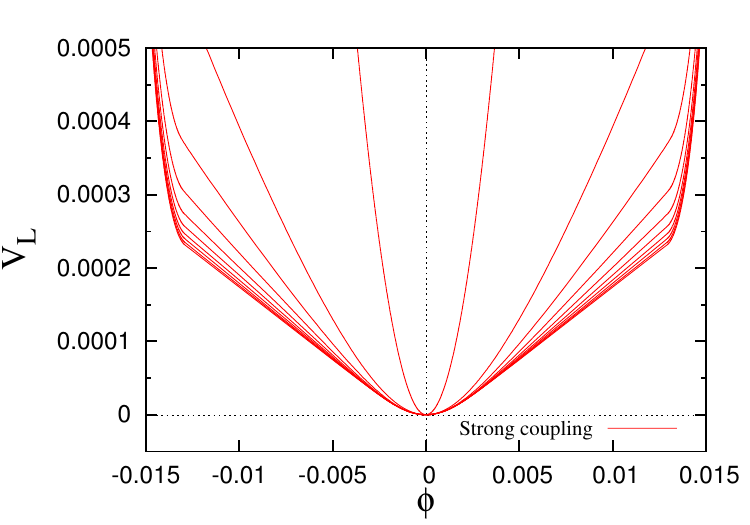}
\caption{Iteration of $V_{\rm L}^{(n)}$ for $g=0.85, \mu=0.7$.}
\label{fig:VLmu}
\end{figure}

We investigate the node length iteration of $M{(n)}$ and $\phi^{(n)}$ for 
$g=0.85, \mu=0.7$ in Fig.\ref{fig:Mmuphi}.
For large enough bare mass $m_0$, the dynamical mass $M$ and $\phi$ are 
generated. However, if we take the
vanishing bare mass limit we have vanishing $M$ and $\phi$.
Then the Legendre effective potential calculated by iteration method 
takes the form depicted in Fig.\ref{fig:VLmu} where we may see the chiral 
symmetry broken points at $\phi \simeq 0.013$, but the minimum of the
effective potential is still at the origin.
Note that the convexity of the Legendre effective potential is 
automatically assured as before.

Referring to another type of analysis of this system, 
we understand that the model with parameter $g=0.85, \mu=0.7$
resides in the chiral symmetry broken phase, that is, the dynamical mass
is generated spontaneously. 
Considering this situation, the node length iteration formulated so far
does not always give correct vacuum for 1st order phase transition case.
In other words, as far as the origin stands for the stable fixed point, 
the vanishing bare mass limit of iteration always goes into the origin.

\begin{figure}[ht]
\centerline{
\FF{60}{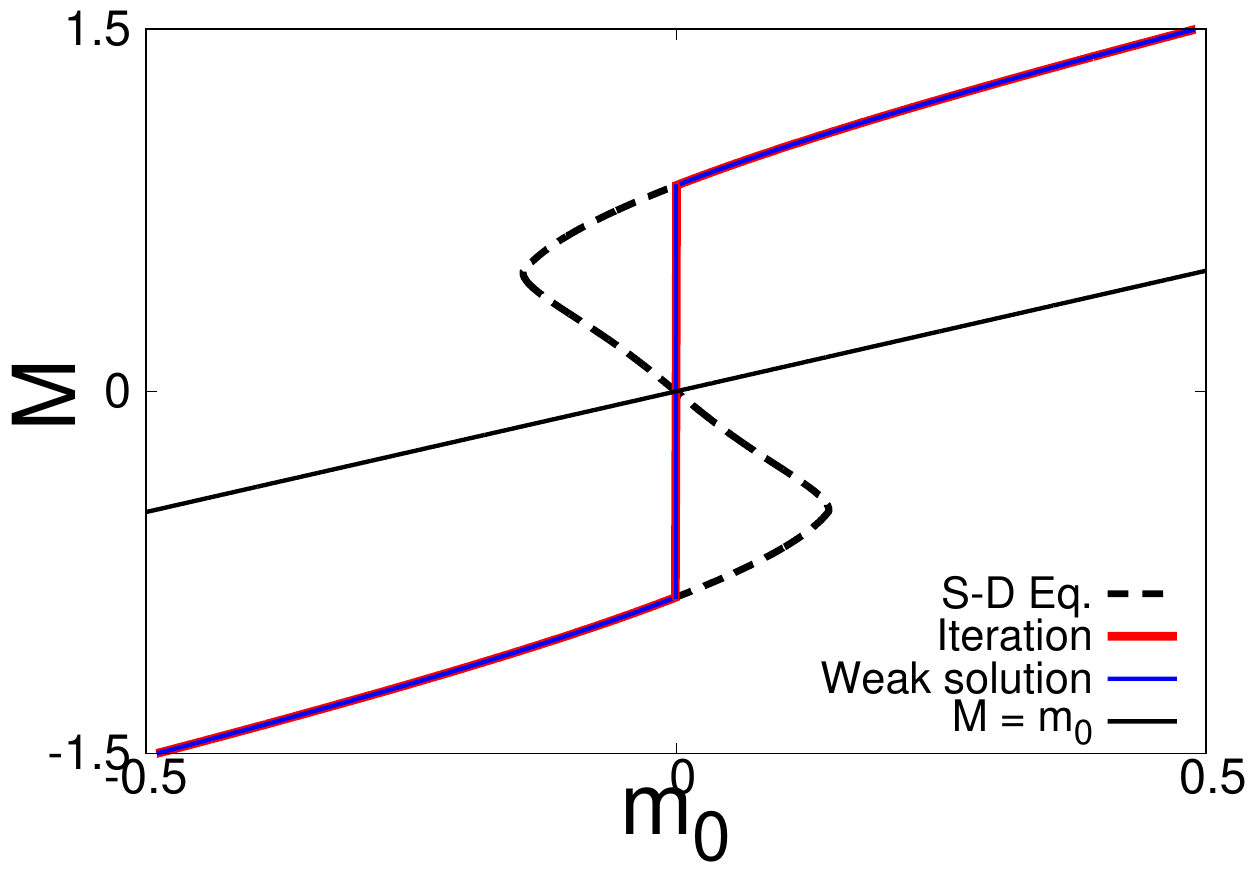}
\FF{60}{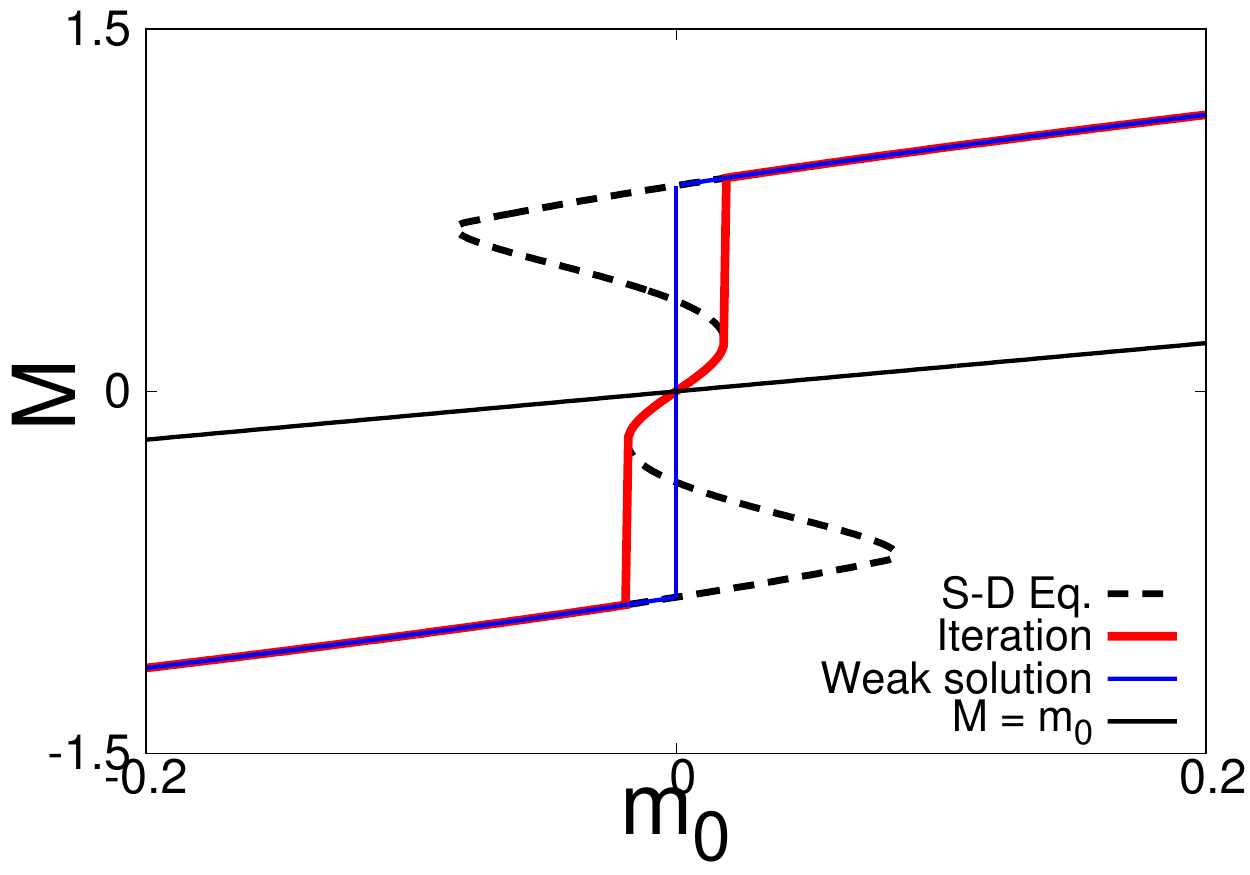}}
\centerline{\hbox to 60mm{\hfil (a) $g=0.85, \mu=0.5$ \hfil}
\hbox to 60mm{\hfil (b) $g=0.85, \mu=0.7$  \hfil}}
\vskip5mm
\centerline{
\FF{60}{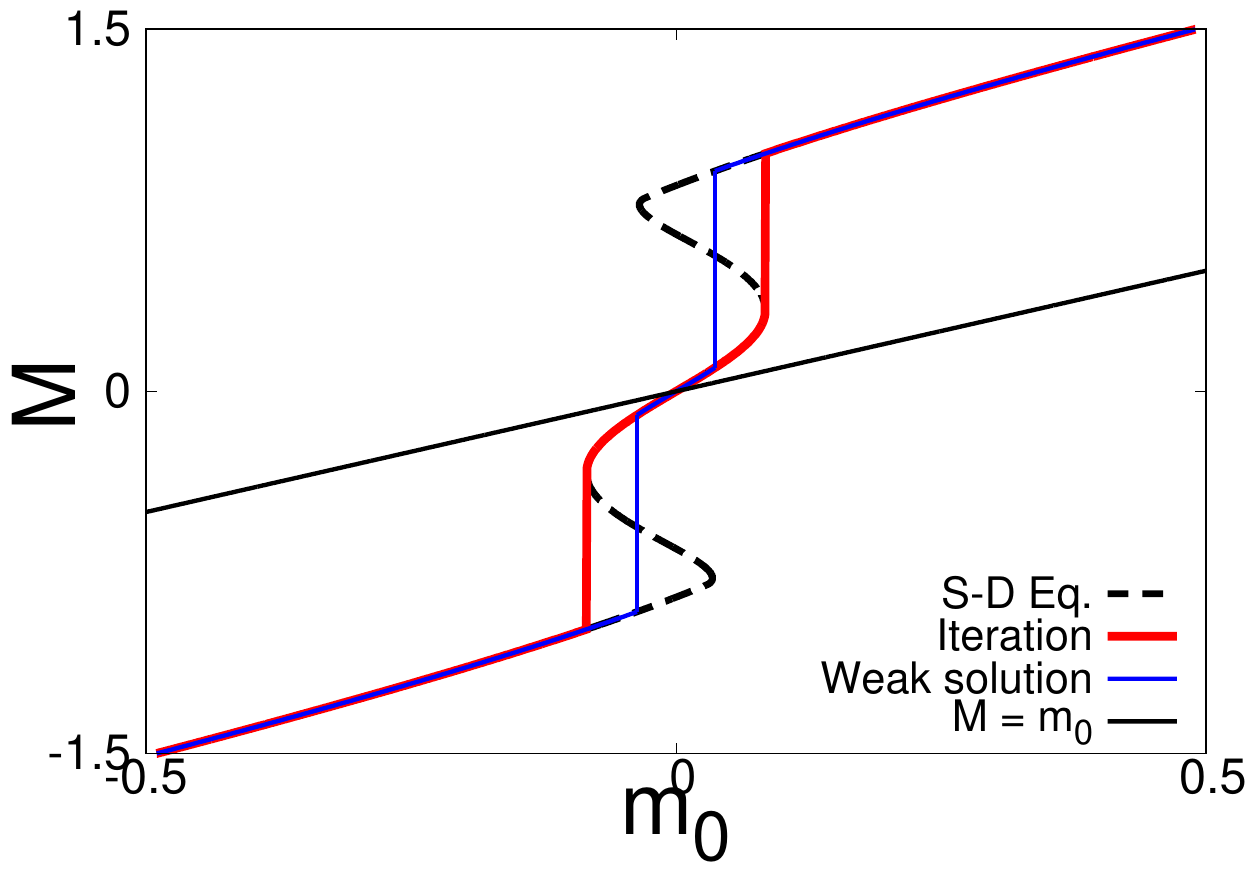}
\FF{60}{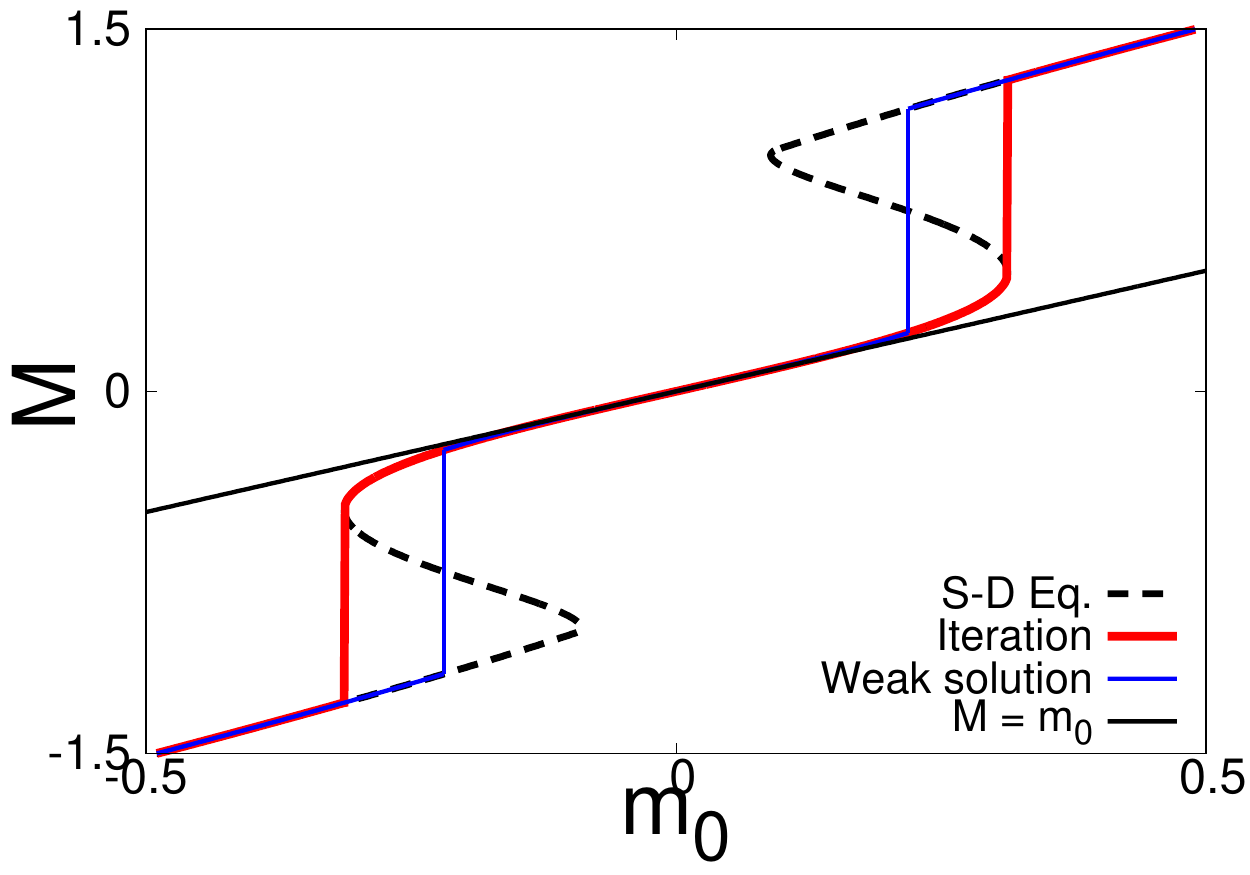}}
\centerline{\hbox to 60mm{\hfil  (c) $g=0.85, \mu=0.8$ \hfil}
\hbox to 60mm{\hfil (d) $g=0.85, \mu=1.0$ \hfil}}
\caption{Comparison of iteration method and weak solution method.}
\label{fig:IteWeakAll}
\end{figure}

Here we prove this property by comparing our results with
those obtained in \cite{Kuma14-1} by using the weak solution 
method of the non-perturbative renormalization group analysis
of the dynamical chiral symmetry breaking. 
In Fig.\ref{fig:IteWeakAll}, we plot the mass as a function of
$m_0$ for various $\mu$ with $g=0.85$.

There are 4 plots in each figure. The dashed line is a multi-valued
function of $m_0$ which is obtained from the Schwinger-Dyson
equation, or the position of all fixed points in our iteration method.
The straight line is $M=m_0$ and shows the initial condition for 
the iteration method.
Starting with this line $M=m_0$, we readily obtain the
infinite $n$ limit of iteration, which is the red curve, just
as drawn in Fig.\ref{fig:Mmuphi}.

The thin blue curve is the weak solution defined in \cite{Kuma14-1}.
The weak solution determines the unique function by a patchwork
of the multi-valued function 
so that the vertical jump line gives the equal area for left and right
sides of the jump. Note that in figure (b), the 4 parts are balanced
totally. This balance of the area assures that the resultant
Legendre effective potential is properly convexified and therefore
the obtained mass $M(m_0)$ is always physically correct.

Therefore, in the parameter regions where the red curve
is different from the blue curve, the iteration method
does not give the physically correct mass.
In figure (a) there is no such region. This is the same type
of multi-valuedness as zero density case, that is the double
well type transition.

In figures (b), (c), (d), the iteration goes wrong at some $m_0$ intervals.
The mismatch is the $m_0$ value where the finite
jump occurs between two stable fixed points.
The blue line separates the multi-valued region into
equal area parts, while the red line always passes the extremum
point.

In figure (b), even the vanishing $m_0$ limit is wrong, 
Note that in cases (c) and (d), although the vanishing 
$m_0$ limit is correct, the difference remain for
large $m_0$ region which is also physical region anyway.

We concentrate on case (b). There is a stable fixed point
around the origin and iteration goes to this fixed point.
It is impossible to give the physically correct result 
(blue line) since this stable fixed point is above the 
initial condition line $M=m_0$ in the positive $m_0$ region near the origin.
To prove this property, 
we expand the transformation function $F(x)$ around the
origin, 
\begin{equation}
F(x) \simeq m_0 + k x,
\end{equation} 
where the stability of the fixed point indicates $k<1$.
Then the fixed point $x^\star_0$,
\begin{equation}
x^\star_0 \simeq \frac{m_0}{1-k},
\end{equation}
is larger than $m_0$.
Therefore the iteration starting with $m_0$ resides in the 
attractive domain of this stable fixed point near the origin,
although this stable fixed point is actually a meta-stable 
state and cannot be the physical vacuum.

This failure of our iteration method in case of 1st order phase 
transition must be reconsidered here from a different view point, 
since it is related to issue of 
multi-valuedness of the infinite sum of Feynman diagrams even within our
method of iteration transformation. In fact, we use the propagator
where the mass is inserted in the denominator, which means we have
done some partial but infinite sum of geometric series first in a particular manner. 

Suppose we set up another shifted iteration system where we add 
fictitious bare mass $M_0$ to the 
propagator and subtract the same quantity in the interaction part. We define
the node length counting so that the four-fermi interactions and the negative 
counter mass interaction play the equal level role. The infinite iteration apparently 
gives the sum of all $1/N$ leading diagrams. 

This shifted iteration, however, 
corresponds exactly to the original iteration starting with a different initial 
point of $M_0+m_0$. 
Then by choosing the fictitious bare mass $M_0$ appropriately, we may select
any fixed point of iteration as the infinite iteration result. This shows that
our iteration method cannot completely avoid the total indefiniteness of the 
infinite sum of $1/N$ leading diagrams. 

\section{Gauge Theory}

\begin{figure}[h!]
 \CFF{60}{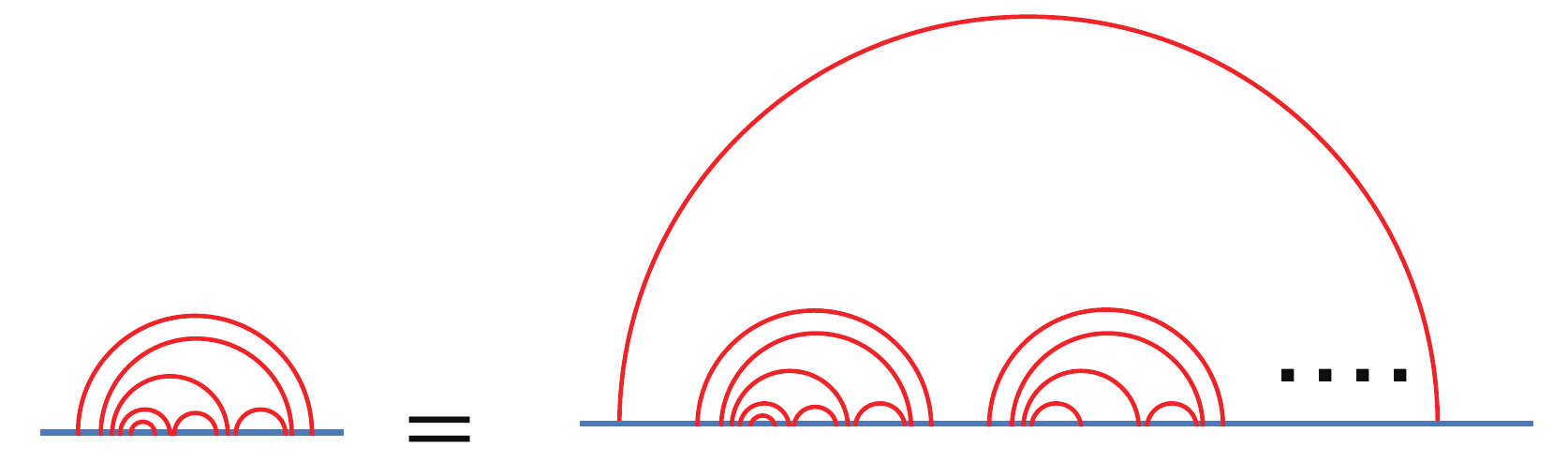}
 \caption{Self-similarity method for ladders.}
 \label{fig:GTSD}
\end{figure}

In this section, we investigate the dynamical chiral symmetry breaking 
in gauge theories. 
We consider the so-called ladder or planar approximation for the fermion self-energy.
The ladder type diagram means that all the gluons are not crossed to each other.
For the total sum of those ladder type diagrams we can find the whole as a part and
set up the self-consistent condition
which is drawn as in Fig.\ref{fig:GTSD}, 
where the straight line represents a fermion and circular lines
represent gluons.
This is an integral equation and there are infinite number of non-trivial
solutions for strong enough gauge coupling constant.

\begin{figure}[h!]
 \CFF{40}{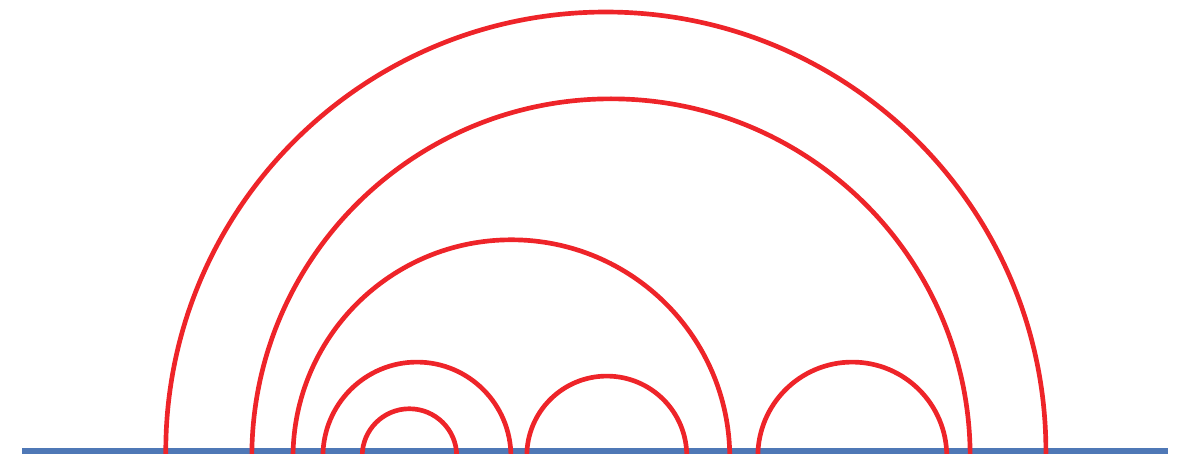}
 \caption{ Example diagram with ladder depth $n=5$.}
 \label{fig:ladderdepth}
\end{figure}

To set up an iterative method to sum up all ladder diagrams, 
we first define the ladder depth for each planar 
diagram. The ladder depth is the maximum number of gluon propagators
counting from the most outer loop towards the fermion propagator. 
For example we show a diagram with ladder depth = 5 in Fig.\ref{fig:ladderdepth}.

\begin{figure}[h!]
 \CFF{80}{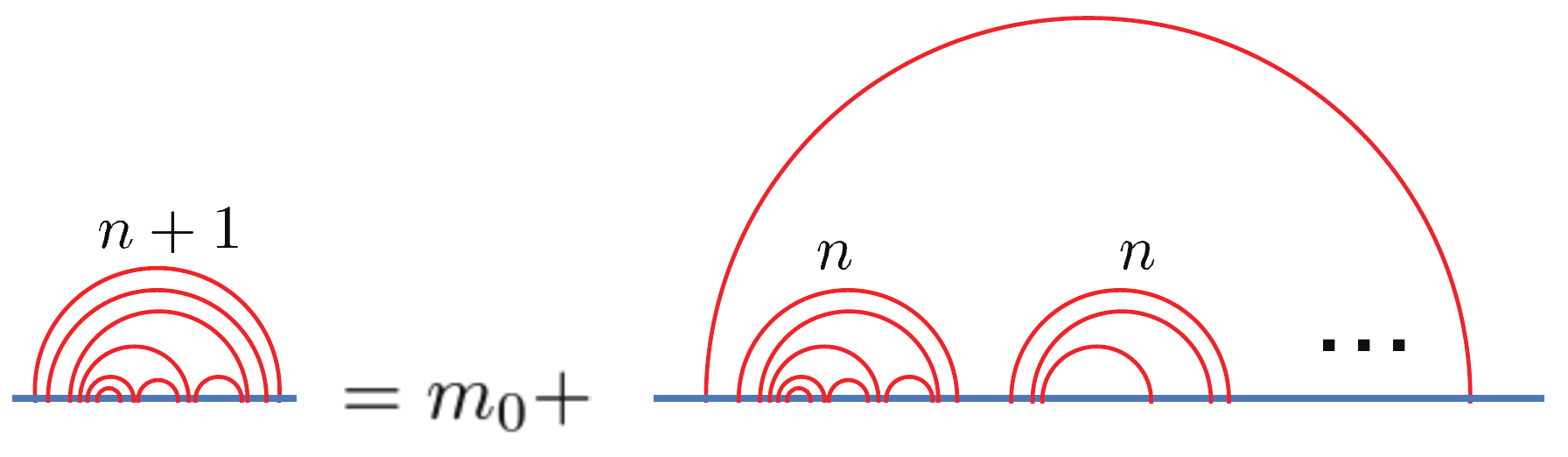}
 \caption{Ladder Depth Iteration}
 \label{fig:LDI}
\end{figure}

Then we define mass function $\Sigma^{(n)}$ 
which contains all planar diagrams whose ladder depth
is no greater than $n$. 
Now we set up the iteration transformation as shown in Fig.\ref{fig:LDI}.

The iteration functional is denoted by
\begin{eqnarray}
 {\Sigma}^{(n+1)}(x)&=&\mathscr{F}\left[{\Sigma}^{(n)}(x)\right], 
\end{eqnarray}
where argument $x$ is the Euclidean momentum squared.
The functional is calculated by one loop integral of 
the Feynman diagram as follows\cite{aoki}:
\begin{eqnarray}
 \mathscr{F}[{\Sigma(x)}]\kern-4mm
&=&\kern -5mm
m_0 +\int^{\Lambda} \frac{d^4 p}{i (2\pi)^4}C_2 (R) \left(\theta (p^2-k^2)g^2 (p^2) +\theta (k^2-p^2) g^2(k^2)\right)\nonumber \nonumber\\
&\quad\times&\frac{g^{\mu\nu} -(p-k)^{\nu}/(p-k)^2 }{ (p-k)^2}\gamma_{\mu}\frac{1}{\Slash{k}-\Sigma(k)}\gamma_{\nu}\nonumber\\
&=&\kern -5mm m_0 + \frac{3C_2(R)}{16\pi^2}\int^{\Lambda^2}_{0}dy \frac{y\Sigma(y)}{y+\Sigma^2} \left(  \frac{g^2(x)}{x} \theta(x-y) + \frac{g^2(y)}{y} \theta(y-x) \right)\nonumber\\
&= &\kern-5mm m_0 +
 \frac{\lambda(x)}{4x}\int_{0}^{x}\frac{y\Sigma(y)dy}{y+\Sigma^2(y)} +
 \int_{x}^{\Lambda^2}\frac{\lambda(y)\Sigma(y)dy}{4(y+\Sigma^2(y))},
\end{eqnarray}
where we set $x=p^2, y=k^2$ and 
$\lambda$ is defined by the running gauge coupling constant $g(x)$, 
\begin{equation}
\lambda(x)= \frac{3g^2(x)}{4\pi^2},
\end{equation}
and $C_2(R)$ is the second Casimir invariant for the fermion representation $R$.

Note that the mass function $\Sigma(x)$ is a function of momentum squared $x$.
The functional $\mathscr{F}$ is now an infinite dimensional map and
there are infinite number of fixed point functions. Our analysis
clarified that only one of fixed point functions is perfectly stable and is
reached by proper initial function $\Sigma^{(0)}(x)=m_0$.

Hereafter numerical calculations are performed for U(1) gauge theory
with fixed gauge coupling constant where $\lambda_{\rm c}=1$.
However, all the results are expected to hold qualitatively for the QCD
with running gauge coupling constant.

\begin{figure}[h!]
 \CFF{100}{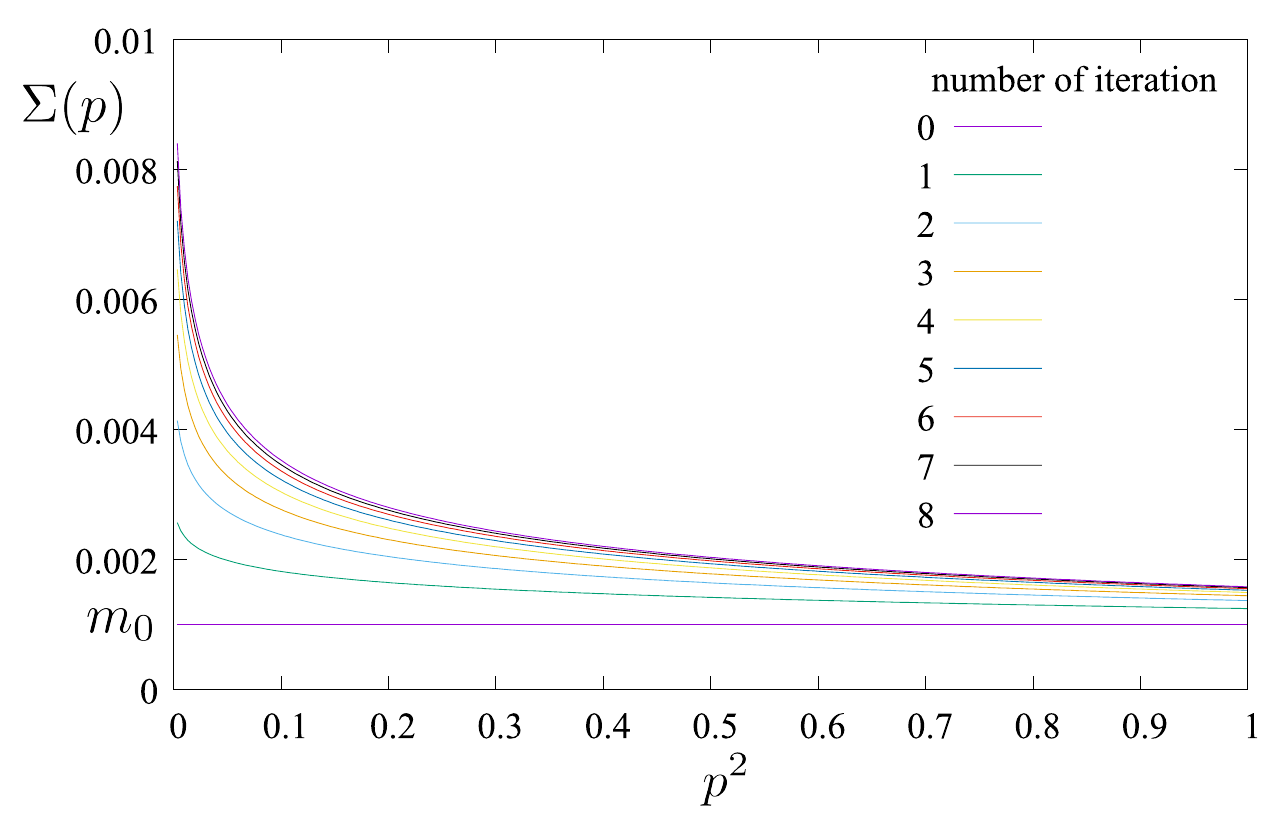}
 \caption{$\Sigma(p^2)$ development due to ladder depth iteration}
 \label{fig:LDIex}
\end{figure}

\begin{figure}[h!]
 \centerline{
 \FF{60}{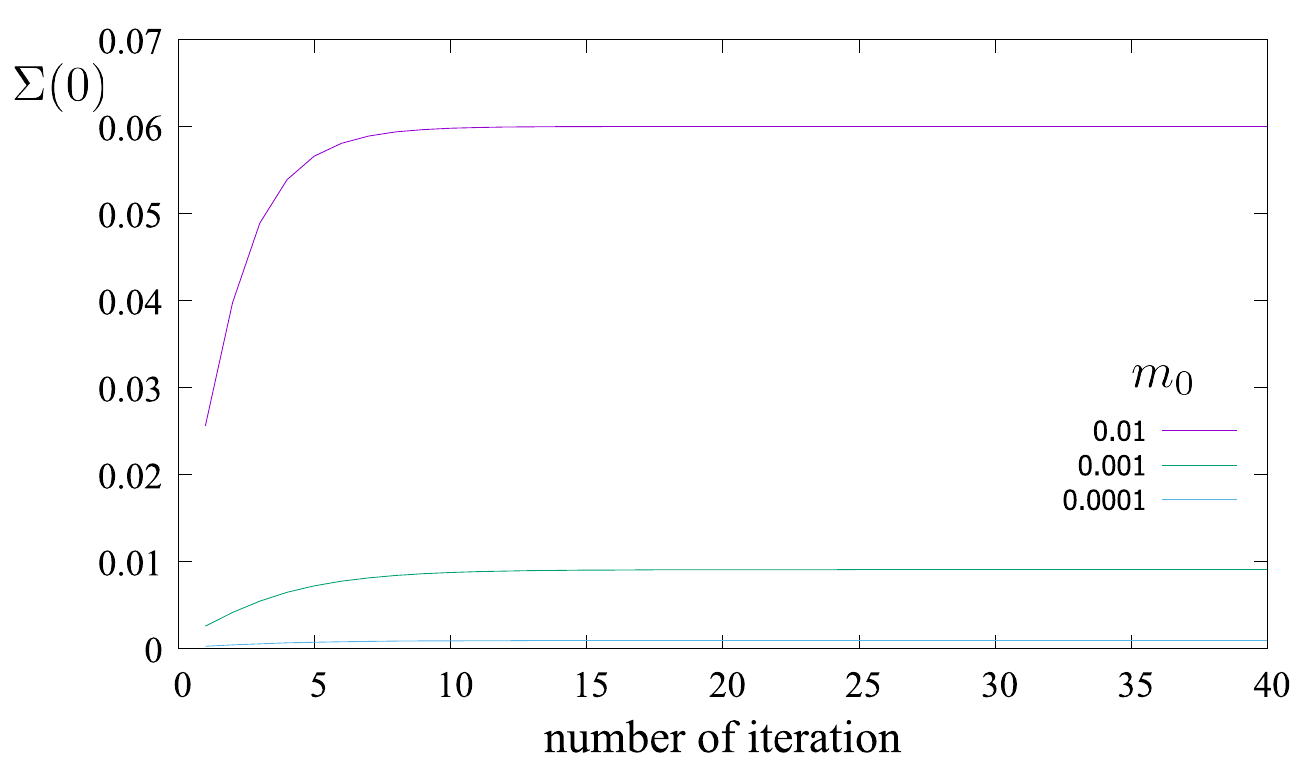}
 \FF{60}{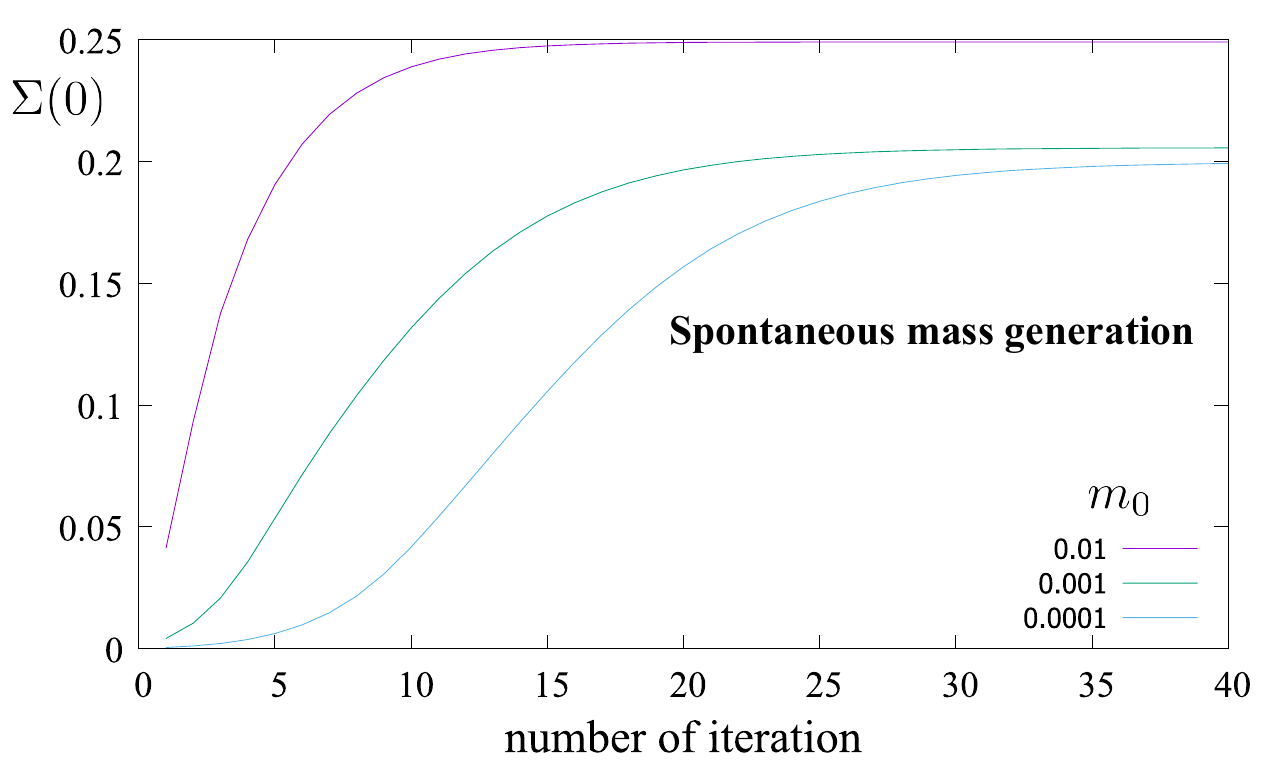}
 }
 \caption{$\Sigma(0)$ iterative behavior and the vanishing bare mass limit.}
 \label{fig:SigmaSW}
\end{figure}

Starting with the initial constant function, $\Sigma^{(n)}(p)$ develops 
according to the ladder depth iteration as shown in Fig.\ref{fig:LDIex}.
The dynamical mass is given by $\Sigma^{(n)}(0)$, and its iterative development
is plotted in Fig,\ref{fig:SigmaSW} for 
$\lambda=0.9$ (left) and $\lambda=1.5$ (right), 
where the left most plot point corresponds
to the bare mass $m_0$.
We decrease $m_0$ in order ($0.01, 0.001, 0.0001$), and 
check the response of $\Sigma^{(n)}(0)$. We can see the switch-on
of the spontaneous mass generation for the strong enough gauge coupling 
constant in Fig.\ref{fig:SigmaSW} (right). 

\begin{figure}[ht]
 \CFF{80}{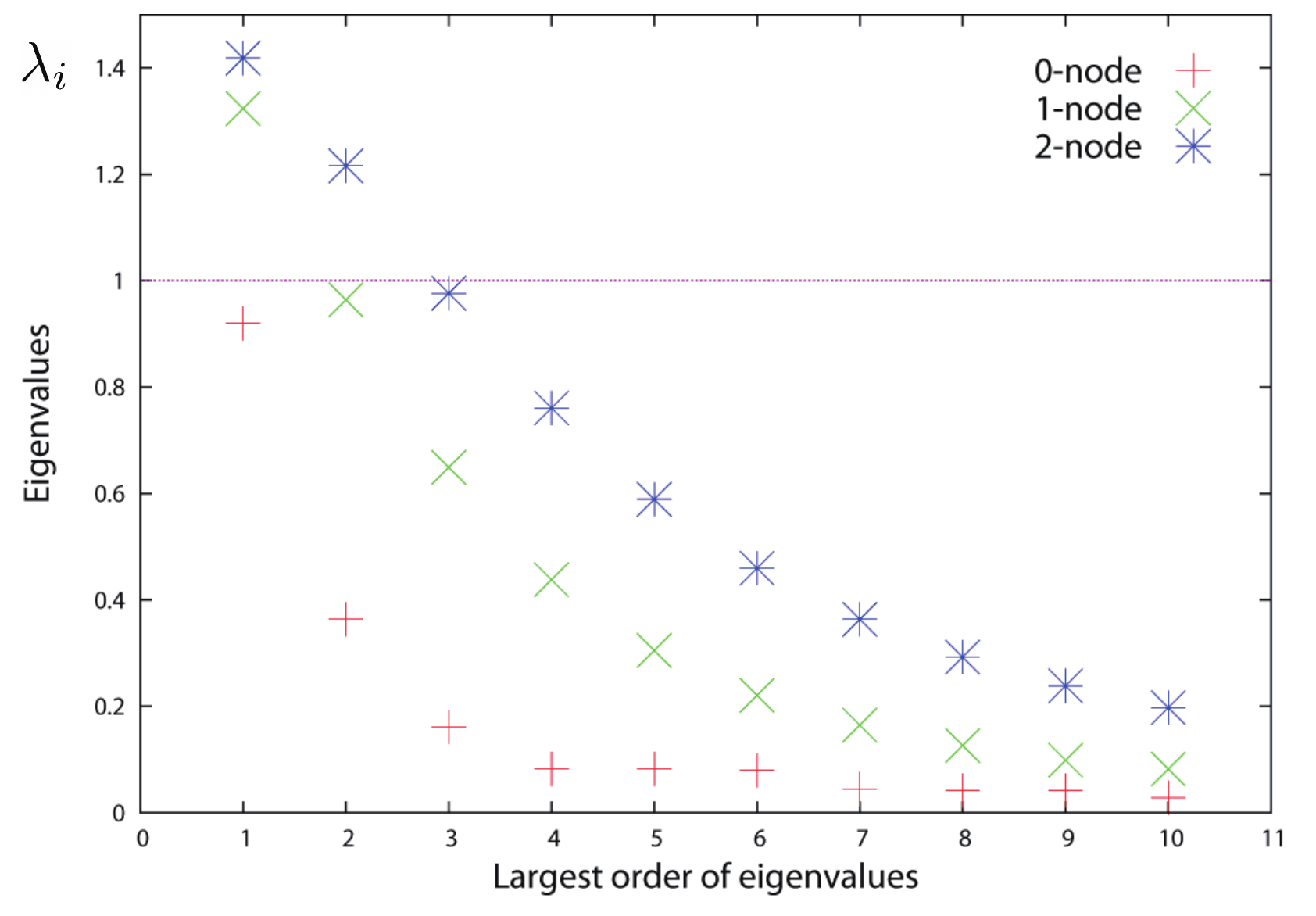}\vskip-0.5cm
 \caption{Several largest eigenvalues for first three fixed point functions}
 \label{fig:EV}
\end{figure}

We check the fixed point of this iterative transformation for the 
super critical ($\lambda>1$) case.
We omit the detailed argument here\cite{SD}. All fixed points are ordered
with respect to the number of nodes in the fixed point function $\Sigma(x)$.
We investigate first three fixed point functions (number of nodes = 0, 1, 2) 
and calculate eigenvalues
around them. We plot eigenvalues in Fig.\ref{fig:EV}.
Eigenvalues less than unity corresponds to attractive direction 
whereas those larger than unity means repulsive direction.

\begin{figure}[h!]
 \CFF{80}{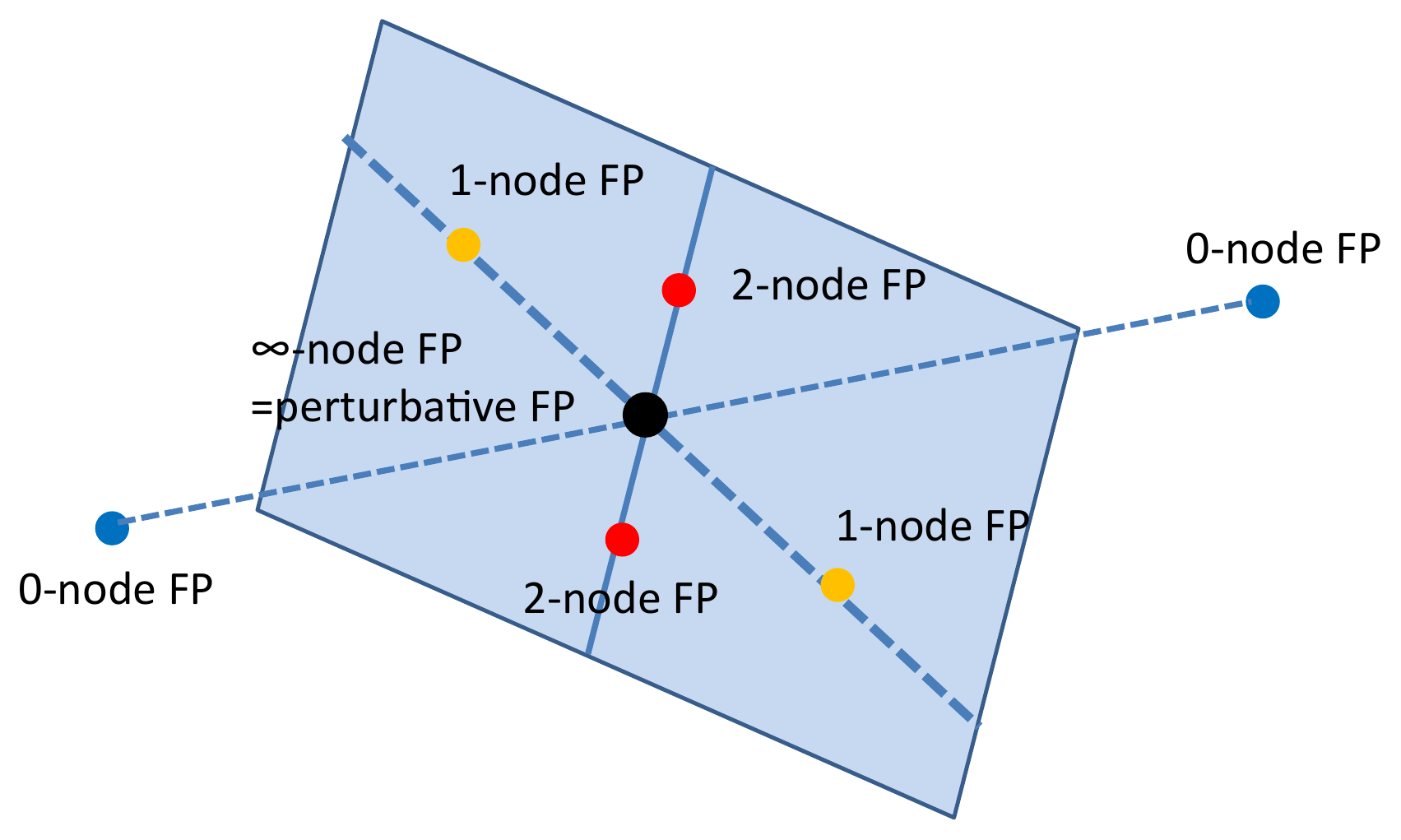} \vskip-0.5cm
 \caption{Hierarchically bifurcating fixed points in the function space}
 \label{fig:PIC}
\end{figure}

As for the 1st fixed point, all eigenvalues are less than unity and
it is a completely attractive fixed point. 
Therefore starting from the initial function $\Sigma^{(0)}=m_0$, we have
a convergent result towards this fixed point function for infinite 
number of iterations.
For the 2nd fixed point there appear one eigenvalue larger than unity
and it is unstable for this direction. In fact, this direction is nothing but 
a route to the 1st fixed point in our function space.
The 3rd fixed point has two eigenvalues larger than unity.
This type of breakup series of fixed points is a very standard image
of the spontaneous symmetry breaking in function space.
It is pictured in Fig.\ref{fig:PIC} where all fixed points are drawn pair-wisely
due to the original chiral symmetry 
(precisely speaking it is U(1) rotational symmetry).

\section{Summary}

We have proposed a new method of calculating the spontaneous
mass generation for the dynamical chiral symmetry breaking.
We work with Nambu--Jona-Lasinio model and gauge theories.
We define the node length classification for NJL mode, and the
ladder depth classification for gauge theories.

The iteration method can directly evaluate the mass without 
any singularity and automatically reaches the physically correct
solution. 
However for the finite density system, where 1st order phase transition
occurs, the iteration method gives a physically inappropriate solution
in case that symmetric vacuum remains meta-stable state.
This miss-match, however, implies a deeper problem in evaluating
infinite sum of Feynman diagrams and give us a subject to be attacked.

\vskip 1cm
We thank illuminating discussions with Yasuhiro Fujii and Masatoshi Yamada.
This work was partially supported by 
JSPS KAKENHI Grant Number 16K13848
and the 2016  Research Grant of Yonago National College of Technology.

\appendix
\section{Geometric series}

To get a clear-cut view of out iterative method, 
we take a simple example of evaluating the geometric series,
\begin{align}
 S=1+r+r^2+r^3+\cdots\ ,
\end{align}
or more definitely, 
\begin{align}
S=\lim_{n\rightarrow \infty} S_n,\ S_n=\sum^n_{k=0}a_k,\ a_k=r^k .
\end{align}

Now we set up the self-consistent equation to evaluate $S$.
Observing the following structure, we see the whole as a part, 
\begin{align}
S=1+\!\!\!\!\!\!\!\underbrace{r^2+r^3+\cdots}_{\displaystyle r(1+r+r^2+\cdots)}\ \hspace{-15pt},
\end{align}
and the self-consistency equation is 
\begin{align}
S=1+rS.
\end{align}
We have a unique solution of it, 
\begin{align}
S=\frac{1}{1-r}\ .
\end{align}
However, at a glance this is not really correct in case of $|r| \geq 1$.
Thus even if there is only one solution, we cannot adopt it always.

We move to another type of method, setting up iterative transformation,
\begin{align}
S_{n+1}&=F(S_n).
\end{align}
For example, the following transformation rule,
\begin{align}
S_{n+1}&=S_n+r^{n+1},
\end{align}
is no good, since the transformation $F$ does depend on $n$.

We find an $n$-independent transformation function,
\begin{align}
S_{n+1}&=r S_n +1, 
\end{align}
which corresponds to 
\begin{align}
F(x)=rx+1.
\end{align}

\begin{figure}[!h]
\CFF{70}{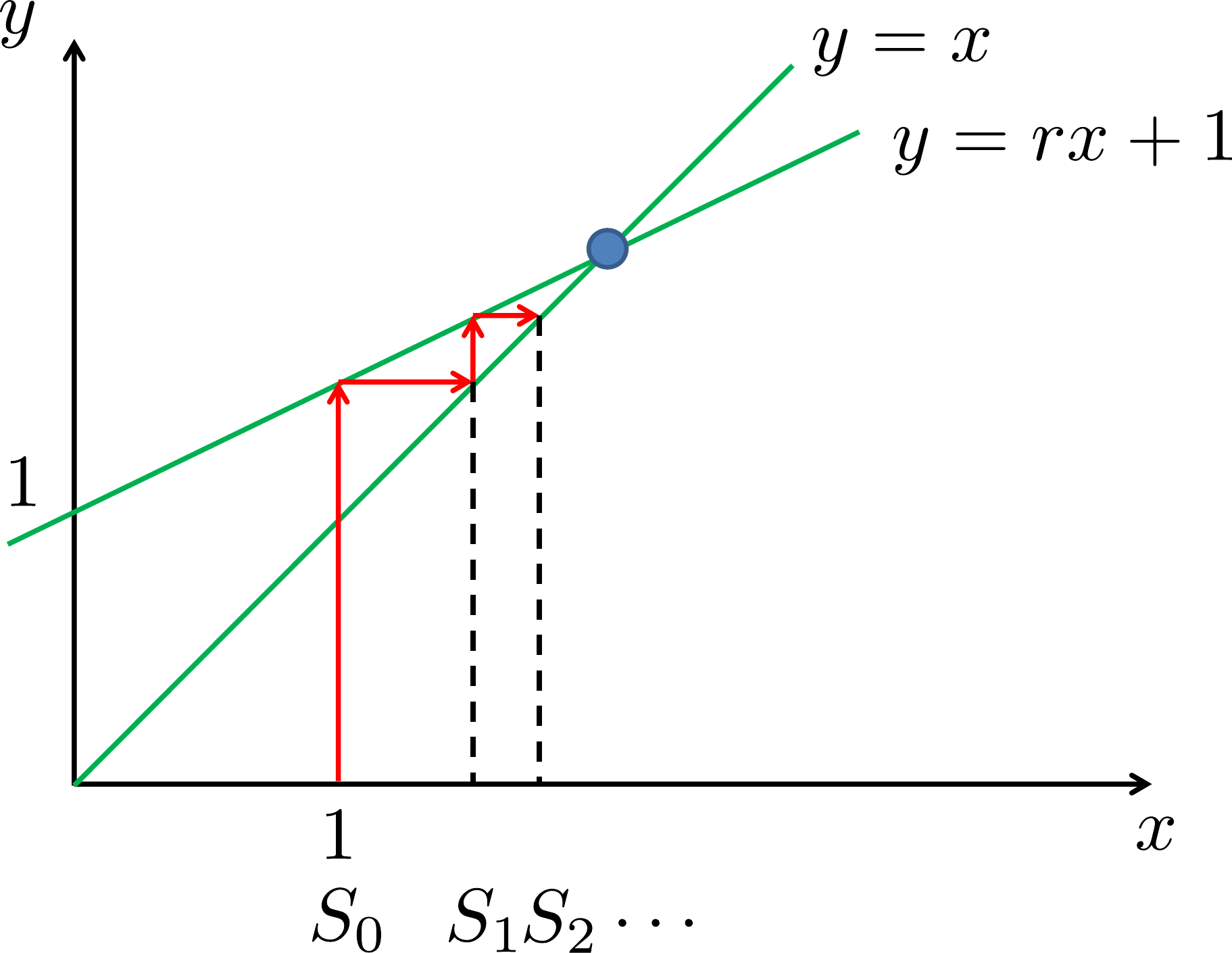}
\caption{Iteration procedure for $|r|<1$.}
\label{fig:GSI}
\end{figure}

This type of iterative transformation is easily solved by a graphical representation.
In Fig.\ref{fig:GSI}, We draw two curves, $y=F(x)$ and $y=x$. 
Starting with the initial value $S_0=1$, we automatically approach to the fixed point: 
\begin{align}
x^*&=F(x^*),
\end{align}
and we have
\begin{equation}
x^*=\frac{1}{1-r},
\end{equation}
which is the result of infinite iterations and gives the correct answer of the sum of geometric series.
Of course, this fixed point coincides with the solution of self-consistent equation before.

\begin{figure}[!h]
\CFF{70}{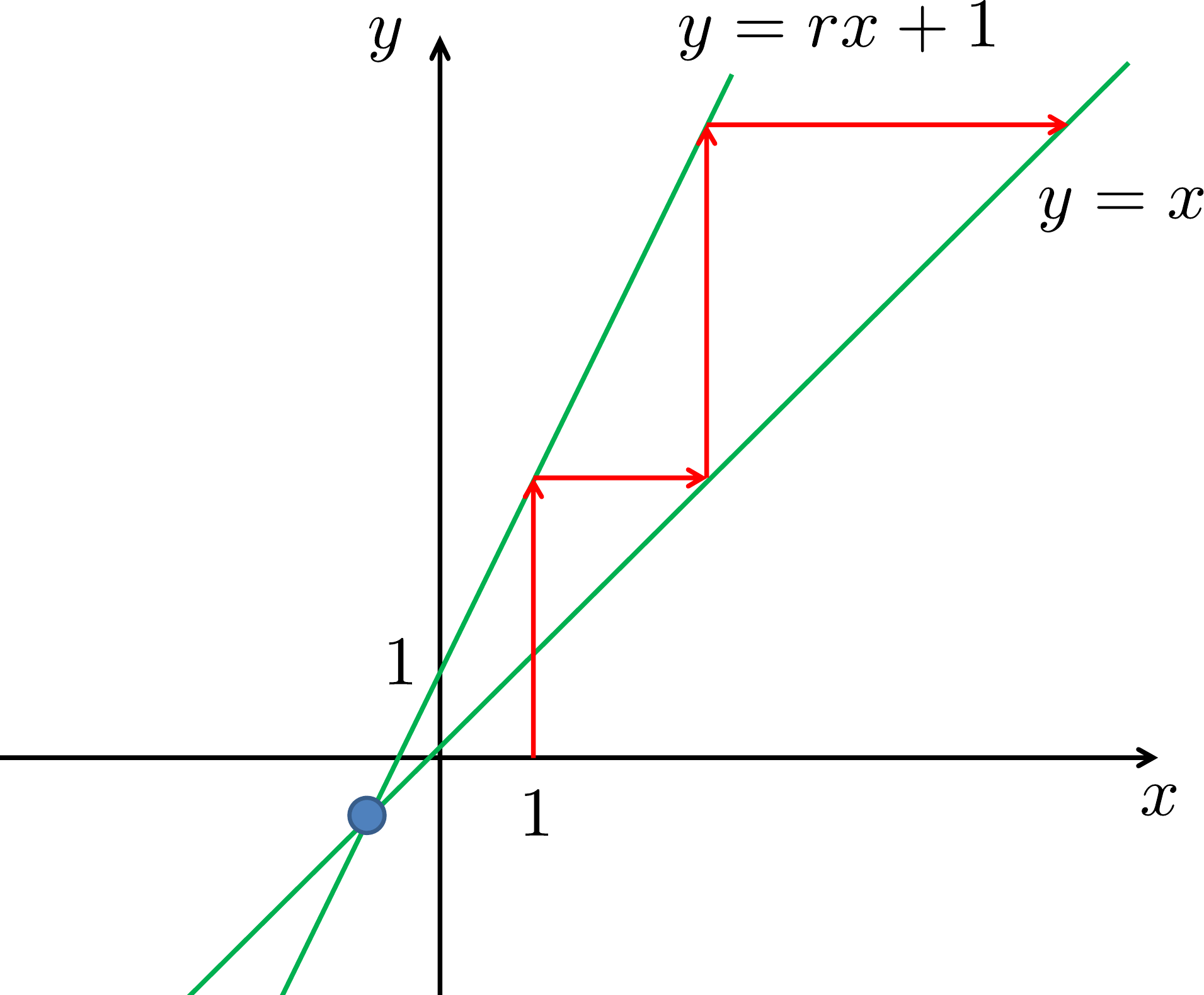}
\caption{Iteration procedure for $|r|>1$.}
\label{fig:GSIb}
\end{figure}

Now we examine what goes on in case $|r|>1$, which is shown in 
Fig.\ref{fig:GSIb}.
We have only one fixed point of the same expression as before.
However, the iteration procedure does not approach to the fixed
point, and instead it separates from the fixed point and diverges towards
infinity.

Thus in this iterative method we successfully evaluate $S$ for any $r$.
The fixed point is exactly the same expression independent of $r$.
However, the eigenvalue around the fixed point is different depending 
on the size of $r$.
For $|r|<1$, the fixed point is an attractor, while for $|r|>1$, it is a repeller.

In this way, we obtain the correct result for $S$ without any additional
inspection. This is the virtue of the iterative method.
Note that in case $|r|<1$, the result after infinite iterations is always 
the same value independent of the initial $S^{(0)}$.
In fact, we can write this another series $\tilde{S}_n$ as
\begin{equation}
\tilde{S}_n = S_n + (\tilde{S}_0 -1)r^n,
\end{equation}
where the difference from the target series is suppressed by $r^n$.

This can be seen as an analogy in the renormalization group analysis of the field theory.
For example, if we add four-fermi interactions to QCD initial Lagrangian, the macro
physics does not change, due to the non-renormalizability of the four-fermi
interactions.

\end{document}